\documentclass{aa}
\usepackage{psfrag}
\usepackage{graphicx}

\begin{document}

\title{Supernovae and the Nature of the Dark Energy}

\author{
  M. Goliath \inst{1,2} \and 
  R. Amanullah \inst{3} \and 
  P. Astier \inst{2} \and
  A. Goobar \inst{3} \and 
  R. Pain \inst{2}
}

\authorrunning{Goliath et al.}

\offprints{M. Goliath}

\institute{
  Swedish Defence Research Agency (FOI), S-172 90 Stockholm, Sweden.
  \email{Martin.Goliath@foi.se}
  \and
  Laboratoire de Physique Nucl\'eaire et de Hautes Energies
  (LPNHE), IN2P3 -- CNRS, Universit\'es Paris VI \& VII, 
  4 Place Jussieu, Tour 33 -- Rez de Chauss\'ee, F-75252 Paris, Cedex
  05, France. \\
  \email{astier@in2p3.fr,rpain@in2p3.fr}
  \and
  Fysikum, Stockholm University, Box 6730, S-113 85 Stockholm, Sweden. \\
  \email{rahman@physto.se,ariel@physto.se}
}

\date{Received ...; accepted ...}

\abstract{
  The use of Type Ia supernovae as calibrated standard candles 
  is one of the most powerful tools to study the expansion history
  of the universe and thereby its energy components. While the analysis
  of some $\sim$50 supernovae at redshifts around $z\sim$0.5 have 
  provided strong evidence for an energy component with negative 
  pressure, ``dark energy'', more data is needed to enable
  an accurate estimate of the amount and nature of this energy.
  This might be accomplished by a dedicated space telescope, 
  the SuperNova / Acceleration Probe (\cite{snap}; SNAP), which aims at 
  collecting a large number of supernovae with $z<2$.
  In this paper we assess the ability of the SNAP mission to determine 
  various properties of the ``dark energy.''
  To exemplify, we expect SNAP, if operated for three years to study
  Type Ia supernovae, to be able  
  to determine the parameters in a linear equation of state
  $w(z)=w_0+w_1\,z$ to within a \emph{statistical} uncertainty of $\pm0.04$ 
  for $w_0$ and
  ${}^{+0.15}_{-0.17}$ for $w_1$ assuming that the universe is known 
  to be flat and an independent high precision
  $(\sigma_{\Omega_m}=0.015)$ measurement of the mass
  density $\Omega_m$, is used to constrain the fit.
  An additional improvement can be obtained if a large number of
  low-$z$, as well as high-$z$, supernovae are included in the sample.   
}

\maketitle

\section{Introduction}

The description of the universe lies at the heart of cosmology, and it
is not surprising that several methods aiming at the determination of
cosmological parameters currently are considered. For example, the power
spectrum of the cosmic microwave background radiation provides means
to determine the total energy content of the universe, for which
recent results of the  balloon-based CMB measurements 
(Jaffe et al. \cite{jaffe}) quote the
value $\Omega_{\rm tot}=1.05\pm0.04$\footnote{This value was derived assuming 
that the Hubble constant is $71\pm8$ km s$^{-1}$Mpc$^{-1}$}. 
Constraints on the matter energy density of the universe, $\Omega_m$, 
can be derived, e.g., from galaxy cluster abundances
(Bahcall \& Fan \cite{bahcall}, Carlberg et al \cite{carlberg}), and 
large-scale structure (Peacock et al \cite{2DF}). These tests
are consistent with $\Omega_m \sim 0.3$, 
see however (Blanchard et al. \cite{blanchard}). 
Furthermore, studies of weak
lensing effects of background objects in mappings of the sky provides
information about the mass distribution in the universe, and thus
measures $\Omega_m$. See, e.g., van Waerbeke et
al. (\cite{waerbeke}) for a discussion of the accuracy of this method.

On top of this, measurements of supernovae at various redshifts
provide a simple way to estimate cosmological parameters
(Goobar \& Perlmutter \cite{gooperl}). In fact, this is the aim of at 
least two collaborations (Riess et al. \cite{high-z}, 
Perlmutter et al. \cite{scp}), both of which recently have
published data in favour of a large energy component attributable to a
cosmological constant, or an evolving scalar field such as
``quintessence'' (Ratra \& Peebles \cite{ratpeeb}, 
Caldwell et al. \cite{caldwell}). The feasibility to determine
the properties of this ``dark energy'' component by using supernova
data has recently been considered by several authors 
(see, e.g., Huterer \& Turner \cite{huturn}, Saini et al. \cite{saini},
Maor et al. \cite{maor}, Astier \cite{astier}, 
Weller \& Albrecht \cite{welalb} and Barger \& Marfatia \cite{barmar}, 
just to list a few),
and conclusions vary significantly. For instance, Huterer
\& Turner (\cite{huturn}), and Saini et al. (\cite{saini}) devise methods
for reconstructing the potential of an acceleration-driving scalar
field, using supernova measurements. On the other hand, Maor et
al. (\cite{maor}) assess the possibility to use supernovae to
distinguish between various cosmological models, allowing for an
evolving equation of state $w(z)$ (which is equivalent to scalar-field
models). They conclude that the prospects for determining the equation
of state in this way are bleak. Barger \& Marfatia (\cite{barmar}) 
support this
latter view, exemplifying how particular data realisations may give
misleading conclusions regarding the dark energy. Again, Weller \&
Albrecht (\cite{welalb}) are more optimistic regarding a determination
of $w(z)$, provided that accurate independent estimates of the matter
energy density $\Omega_m$ are at hand.
As already emphasized by one of us (Astier \cite{astier}), much of the
discrepancies stem from differences in the initial assumptions, e.g.,
in the prior knowledge of $\Omega_m$. 

In this paper we intend to study the extent to which properties of the
dark energy can be determined, assuming that observations of a large number
of supernovae at high redshifts become available. Such data could be
provided by the projected SNAP satellite mission. In section
\ref{sec:dl} we establish our notation and give the expression for the 
luminosity distance $d_L$. 
Section \ref{sec:snap} contains investigations of
different scenarios in line with the SNAP proposal
(\cite{snap}). Confidence regions for cosmological parameters are
obtained for various situations. Section \ref{sec:add} considers
the relative importance of events at various redshifts by
investigating the effect of adding a small sample at various 
specific redshifts. In section \ref{sec:lens}, we
analyse the systematic errors in cosmological parameter estimation
that are caused by gravitational lensing. We end with a
discussion of the main conclusions in section
\ref{sec:discuss}. Appendix \ref{app:method} outlines the construction
of our log-likelihood functions in some detail.

\section{Apparent magnitude and luminosity distance}\label{sec:dl}

We intend to investigate the feasibility to determine cosmological
parameters $\vec{\theta}=(\Omega_m,\Omega_X,w_0,w_1)$ by using observational
data from supernovae at different redshifts $z$. Here, $\Omega_m$ and
$\Omega_X$ denote the present-day energy density parameters of ordinary
matter $\Omega_m(z)$ and a ``dark energy'' component $\Omega_X(z)$, 
respectively. The equation of
state $w(z)$ of the dark energy is parametrised by $(w_0,w_1)$ to
linear order: $w(z)\approx w_0+w_1\,z$.

The apparent magnitude $m$ of a supernova at redshift $z$, assuming the
cosmology $\vec{\theta}$, is given by
\begin{eqnarray}
    m(\vec{\theta},{\cal M},z)&=&
    {\cal M}+5\log_{10}\left[d'_L(\vec{\theta},z)\right],\\
    {\cal M}&=&25+M+5\log_{10}(c/H_0),\label{eq:Mscript}
\end{eqnarray}
where $M$ is the absolute magnitude of the supernova, and 
$d'_L\equiv H_0\,d_L$ is the \emph{$H_0$-independent} luminosity distance, 
where $H_0$ is the Hubble parameter\footnote{In the expression for 
${\cal M}$, the units of $c$ and $H_0$ are km s$^{-1}$ and 
km s$^{-1}$ Mpc$^{-1}$, respectively.}. 
Hence, the intercept ${\cal M}$ contains the ``unwanted''
parameters $M$ and $H_0$ that apply equally to all magnitude
measurements (we do not consider evolutionary effects $M=M(z)$).
The $H_0$-independent luminosity distance $d'_L$ is given by
\begin{eqnarray}
    d'_L&=&\left\{
    \begin{array}{ll}
      (1+z)\frac{1}{\sqrt{-\Omega_k}}\sin(\sqrt{-\Omega_k}\,I) , &
      \Omega_k<0\\
      (1+z)\,I , & \Omega_k=0\\
      (1+z)\frac{1}{\sqrt{\Omega_k}}\sinh(\sqrt{\Omega_k}\,I) , &
      \Omega_k>0\\
    \end{array}
    \right. \\
    \Omega_k&=&1-\Omega_m-\Omega_X,\\
    I&=&\int_0^z\,\frac{dz'}{H'(z')} ,\\
    H'(z)&=&H(z)/H_0=\nonumber\\
    &&\sqrt{(1+z)^3\,\Omega_{m}+
    f(z)\,\Omega_{X}+(1+z)^2\,\Omega_{k}}, \\
    f(z)&=&\exp\left[3\int_0^z\,dz'\,\frac{1+w(z')}{1+z'}\right] ,
\end{eqnarray}
where we consider an equation of state linear in $z$:
\begin{equation}
  w(z)=w_0+w_1\,z.
\end{equation}

\section{Statistical uncertainties for one year of SNAP data}\label{sec:snap}

The SuperNova/Acceleration Probe (\cite{snap}; SNAP) is a proposed
two-meter satellite telescope specifically designed to discover and
follow supernovae over a wide redshift range. In particular, such an
instrument would be able to provide photometry and spectra of more than
2000 SN Ia per year (SNAP proposal \cite{snap}). We will investigate the 
accuracy of
cosmological parameter estimations based on one year of SNAP data. 
To this end, we assume that 2000 supernovae are obtained in the
redshift interval $z\in[0,1.2]$, and an additional 100 at high
redshift $z\in[1.2,1.7]$. The individual measurement precision is
assumed to be $\Delta m=0.15$ magnitudes, including the intrinsic spread of 
supernova brightnesses. 
We divide the redshift interval
into bins of equal size $\Delta z=0.05$. In summary:
\begin{center}
  \begin{tabular}{c@{\quad}c@{\quad}c@{\quad}c@{\quad}c}
    $z$ range & \# SNe & \# bins & \# SNe/bin & prec./bin \\
    & & & & {[mag.]} \\ \hline
    $[0.0, 1.2]$ & 2000 & 24 & 83.33 & 0.0164 \\
    $[1.2, 1.7]$ &  100 & 10 & 10    & 0.0474 \\
  \end{tabular}
\end{center}
We will use the fiducial cosmology from the SNAP proposal: 
$\vec{\theta}_{\rm true}=(0.28,0.72,-1,0)$. 
These assumptions adhere to the SNAP proposal (\cite{snap}), except that
we do not include any systematic errors. However, in section 
\ref{sec:lens} we will investigate the effects of gravitational
lensing on cosmological parameter estimations.

Below, we consider several different scenarios and present confidence 
regions for parameter estimates. The methodology that has been employed is 
outlined in App. \ref{app:method}. The one-parameter one-sigma uncertainties 
for the various cases are summarised in tables 
\ref{tab:sigma1} -- \ref{tab:sigma3}.

\subsection{Confidence regions for $(\Omega_m,\Omega_X)$}

\begin{figure}
  \centering
  \includegraphics[width=\hsize]{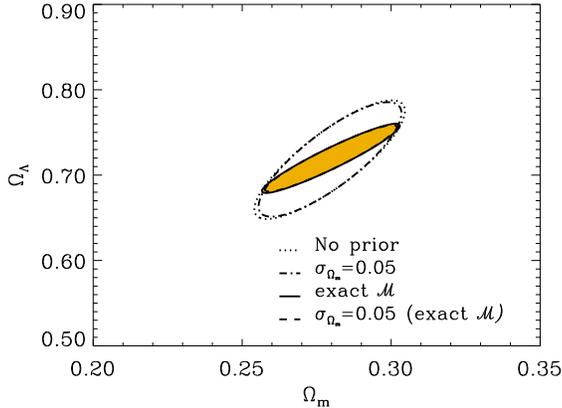}
  \caption{68.3 \% confidence regions for
    $(\Omega_m,\Omega_X)$ in the one-year SNAP scenario. 
    The filled region assumes exact knowledge of ${\cal M}$
    (solid and  dashed lines approximately coincide). 
    A full three-parameter fit with no prior knowledge of
    ${\cal M}$ is assumed for the two larger confidence regions: 
    the region with a dotted line assumes no prior knowledge of $\Omega_m$,
    while the dash-dotted line  assumes a prior knowledge with $\Omega_m$
    Gaussian around the true value and
    $\sigma_{\Omega_m{\rm-prior}}=0.05$.}
  \label{fig:snap01} 
\end{figure}

First, let us assume that it is known that the dark energy corresponds
to a cosmological constant, so that $(w_0,w_1)=(-1,0)$. In this
particular case, $\Omega_X$ is often denoted $\Omega_\Lambda$. Figure
\ref{fig:snap01} shows confidence regions for $(\Omega_m,\Omega_X)$
for various situations. As regards $\Omega_m$, we assume either no
prior knowledge, or else prior knowledge with $\Omega_m$ Gaussian
around the true value with $\sigma_{\Omega_m{\rm-prior}}=0.05$. Concerning the
intercept ${\cal M}$, we assume either exact knowledge of ${\cal M}$, or
no prior knowledge at all. The latter case involves the expression
$\chi^2_{{\cal M}{\rm-int}}$, given in appendix
\ref{app:Mscript-int}.

\begin{figure}
  \centering
  \includegraphics[width=\hsize]{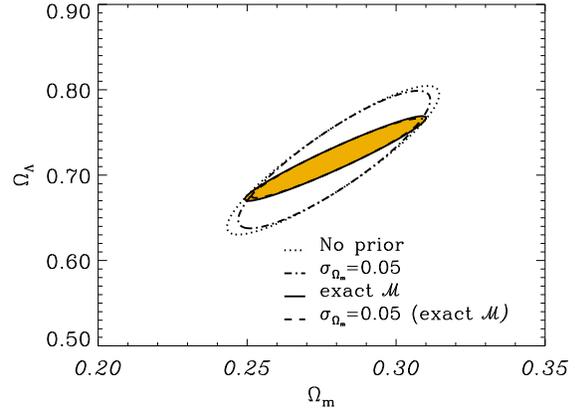}
  \caption{68.3 \% and 95\% confidence regions for
    $(\Omega_m,\Omega_X)$ in the one-year SNAP scenario without the 100 events
    for which $z\in[1.2,1.7]$.
    The filled region (solid line) assumes exact knowledge of ${\cal M}$,
    and the dashed line within the filled region assumes also a prior 
    knowledge with $\Omega_m$ Gaussian around the true value and 
    $\sigma_{\Omega_m{\rm-prior}}=0.05$.
    A full three-parameter fit with no prior knowledge of
    ${\cal M}$ is assumed for the two larger confidence regions: 
    the region with a dotted line assumes no prior knowledge of $\Omega_m$,
    while the dash-dotted line  assumes a prior knowledge with $\Omega_m$
    Gaussian around the true value and $\sigma_{\Omega_m{\rm-prior}}=0.05$.}
  \label{fig:snap19} 
\end{figure}

\begin{figure}
  \centering
  \includegraphics[width=\hsize]{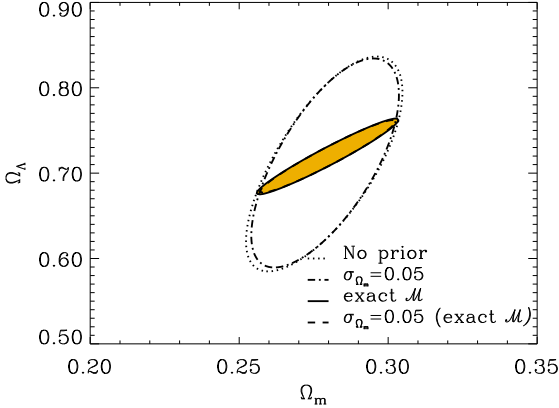}
  \caption{68.3 \% and 95 \% confidence regions for
    $(\Omega_m,\Omega_X)$ in the one-year SNAP scenario with a constant rate
    per co-moving volume for $z\in[0,1.2]$, and the 100 $z\in[1.2,1.7]$ events
    uniformly distributed.
    The filled region assumes exact knowledge of ${\cal M}$ 
    (solid and  dashed lines approximately coincide).
    A full three-parameter fit with no prior knowledge of
    ${\cal M}$ is assumed for the two larger confidence regions:
    the region with a dotted line assumes no prior knowledge of $\Omega_m$,
    while the dash-dotted line  assumes a prior knowledge with $\Omega_m$
    Gaussian around the true value and
    $\sigma_{\Omega_m{\rm-prior}}=0.05$.}
  \label{fig:snap24} 
\end{figure}

Under the assumption of exact knowledge of ${\cal M}$, we find the
uncertainties 
in $\Omega_m$ and $\Omega_X$ to be $\sigma_{\Omega_m}\approx0.015$,
$\sigma_{\Omega_X}\approx0.027$. However, with no prior knowledge of
${\cal M}$, the uncertainty in $\Omega_X$ grows almost by a factor of
two. Note that the uncertainty in $\Omega_m$ is essentially
unaffected. Hence, imposing the prior knowledge of $\Omega_m$ as
outlined above, does not significantly affect the size of the confidence 
region.  
To emphasize the importance of obtaining at least a few supernovae at
high redshift, we perform the same calculation including only the
events for which $z<1.2$, see figure \ref{fig:snap19}. Even
though there were only 100 such supernovae in the original calculation,
they result in about 25 \% better determination of $\Omega_X$. Thus,
it seems to be well-worth the effort to devise a scheme for obtaining
these high-$z$ events. On the other hand, in order to reduce the
sensitivity to uncertainty in the intercept ${\cal M}$, it is
important to have supernovae at {\it low} redshifts. To illustrate
this, we have examined a situation where the redshifts of the 2000
supernovae at $z\in[0,1.2]$ are distributed according to a constant
rate per co-moving volume element (as opposed to the uniform distribution used
before). The few events for $z\in[1.2,1.7]$ are still considered to be
uniformly distributed. As seen in figure \ref{fig:snap24}, this does
hardly affect the uncertainties when ${\cal M}$ is exactly
known. However, for the worst-case scenario of no prior knowledge of
${\cal M}$, the uncertainty in $\Omega_X$ grows almost by a factor of
three. The relative importance of events at various redshifts is further
discussed in section \ref{sec:add}.

\subsection{Confidence regions for $(\Omega_m,w_0)$ or $(\Omega_X,w_0)$}

\begin{figure}
  \centering
  \includegraphics[width=\hsize]{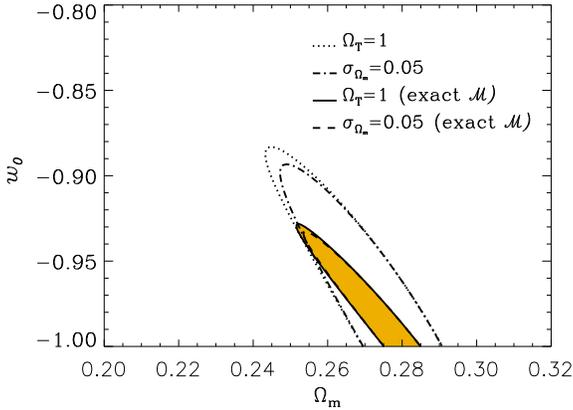}
  \caption{68.3 \%  confidence regions for
    $(\Omega_m,w_0)$ in the one-year SNAP scenario with the flatness
    assumption $\Omega_{\rm tot}=1$. 
    The filled region (solid line) assumes exact knowledge of ${\cal M}$, 
    the dashed line within the filled region assumes also a prior knowledge 
    with $\Omega_m$ Gaussian around the true value and 
    $\sigma_{\Omega_m{\rm-prior}}=0.05$.
    The dotted and dash-dotted lines assume no prior knowledge of 
    ${\cal M}$, without and with $\Omega_m$ prior, respectively.}
    \label{fig:snap02} 
\end{figure}

\begin{figure}
  \centering
  \includegraphics[width=\hsize]{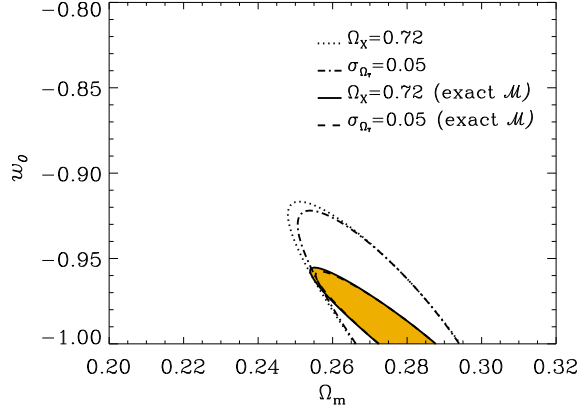}
  \caption{68.3 \% confidence regions for
    $(\Omega_m,w_0)$ in the one-year SNAP scenario assuming exact knowledge of
    $\Omega_X$, i.e., {\em no} prior knowledge on the geometry. 
    The filled region (solid line) assumes exact knowledge of ${\cal M}$, 
    the dashed line within the filled region assumes also a prior knowledge 
    with $\Omega_m$ Gaussian around the true value and 
    $\sigma_{\Omega_m{\rm-prior}}=0.05$.
    The dotted and dash-dotted lines assume no prior knowledge of 
    ${\cal M}$, without and with $\Omega_m$ prior, respectively.}
  \label{fig:snap03} 
\end{figure}

\begin{figure}
  \centering
  \includegraphics[width=\hsize]{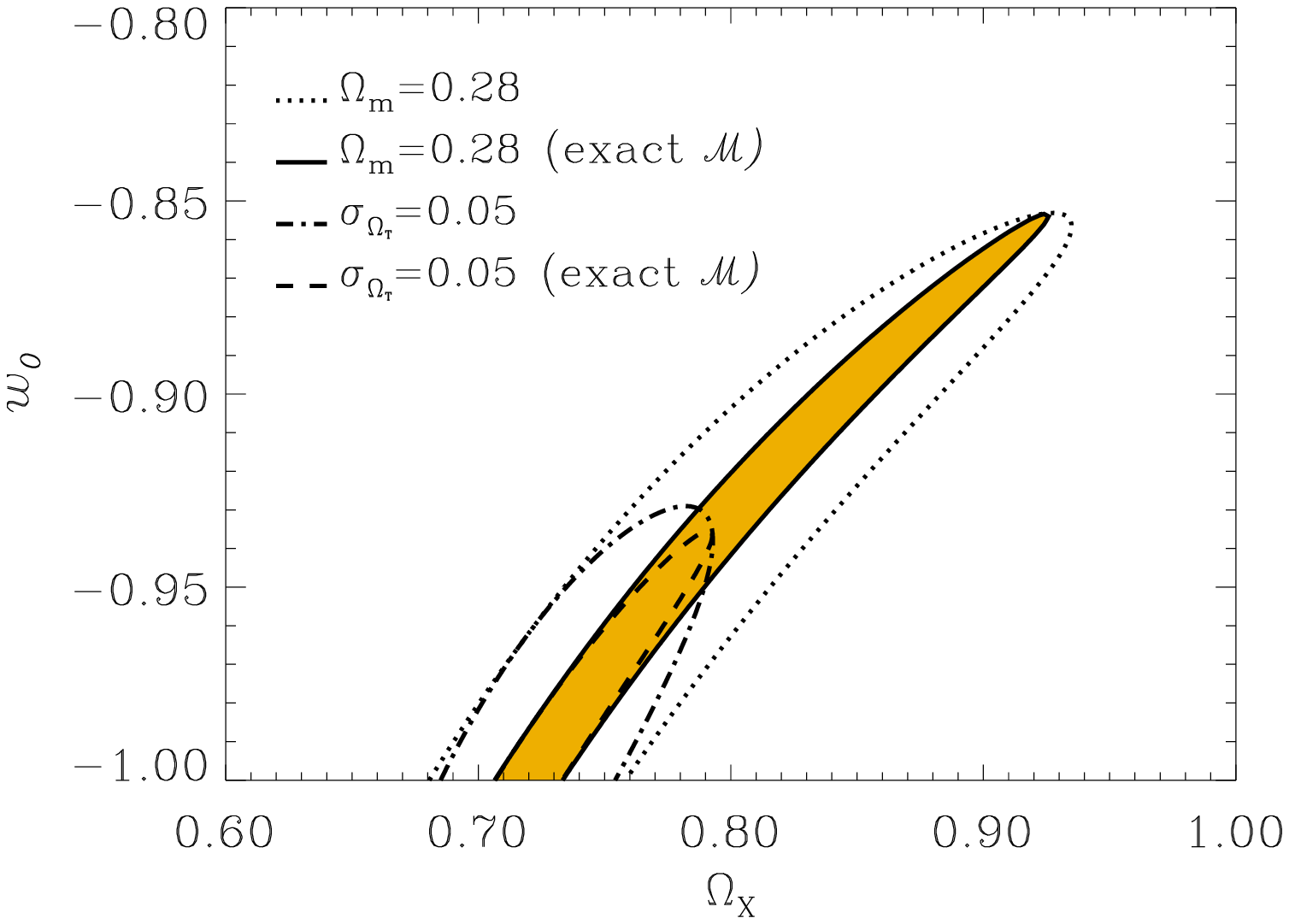}
  \caption{68.3 \%  confidence regions for
    $(\Omega_X,w_0)$ in the one-year SNAP scenario assuming exact knowledge of
    $\Omega_m$. 
    The filled region (solid line) assumes exact knowledge of ${\cal M}$, 
    the dashed line within the filled region assumes also a prior knowledge 
    with $\Omega_{\rm tot}$ Gaussian around the true value and
    $\sigma_{\Omega_{\rm tot}{\rm-prior}}=0.05$.
    The dotted and dash-dotted lines assume no prior knowledge of 
    ${\cal M}$, without and with $\Omega_{\rm tot}$ prior, respectively.}
  \label{fig:snap04} 
\end{figure}

Next, we assume that the equation of state of the dark energy can be
described by a constant $w=w_0$, so that $w_1=0$. Figures
\ref{fig:snap02} -- \ref{fig:snap04} show confidence regions for
$(\Omega_m,w_0)$ or $(\Omega_X,w_0)$ under different assumptions that
fix one parameter in the expression
$\Omega_{\rm tot}=\Omega_m+\Omega_X$: figure \ref{fig:snap02} fixes the
total energy density $\Omega_{\rm tot}=1$, which corresponds
to a flat universe; figure \ref{fig:snap03} assumes that the density of
the dark energy is known exactly, $\Omega_X=0.72$; figure
\ref{fig:snap04} assumes that the energy density of ordinary matter is
exactly known, $\Omega_m=0.28$.
As before, we consider either exact knowledge of ${\cal M}$, or no
prior knowledge. We also consider prior knowledge of either $\Omega_m$
or $\Omega_{\rm tot}$, with spread $\sigma_{\rm prior}=0.05$.
It turns out that the (unrealistic) case where $\Omega_X$ is well-known
gives the best determination of the other parameters under
consideration, and that a good knowledge of $\Omega_{\rm tot}$ is
preferred over a well-determined $\Omega_m$. 

\begin{figure}
  \centering
  \includegraphics[width=0.49\hsize]{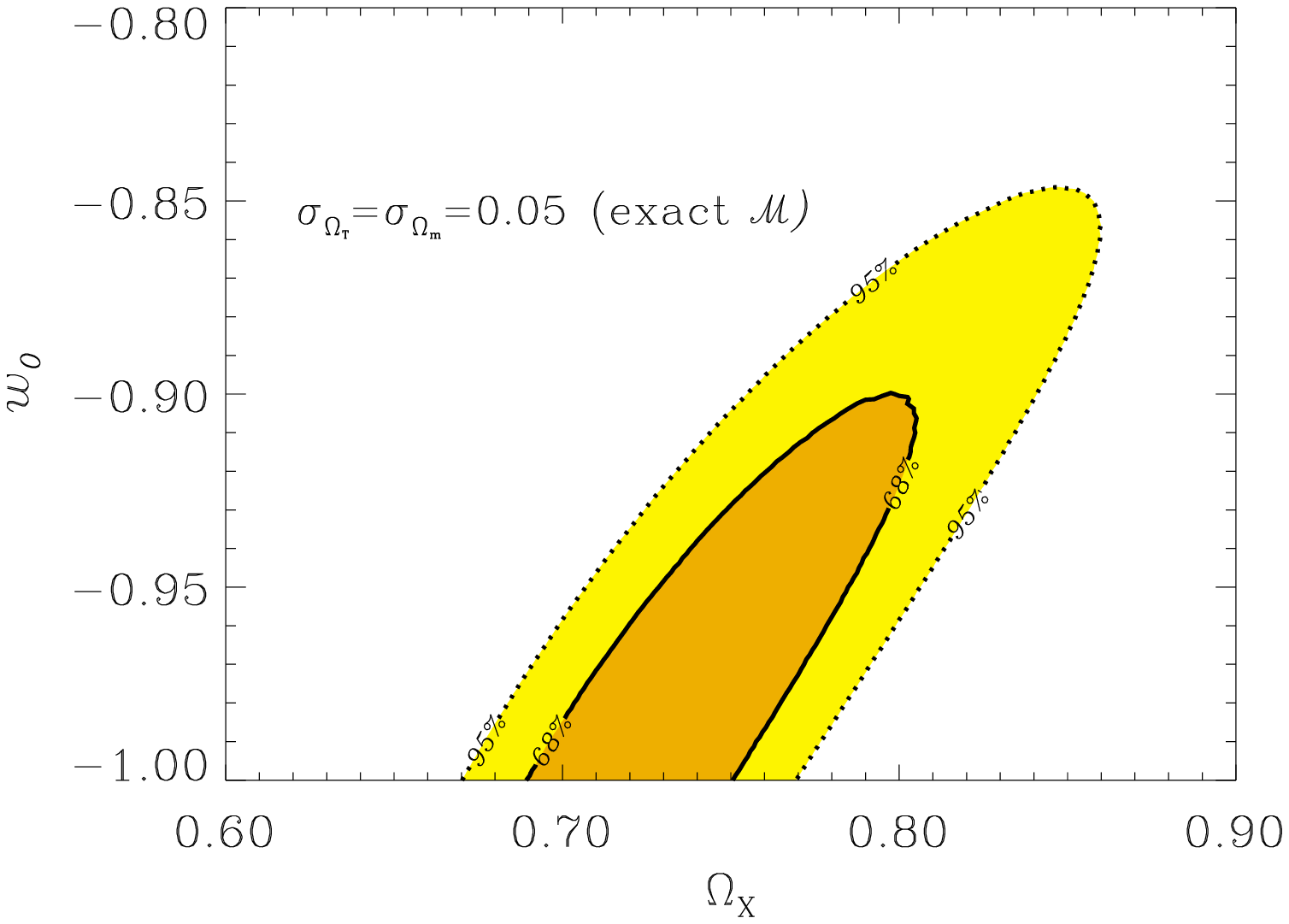}
  \includegraphics[width=0.49\hsize]{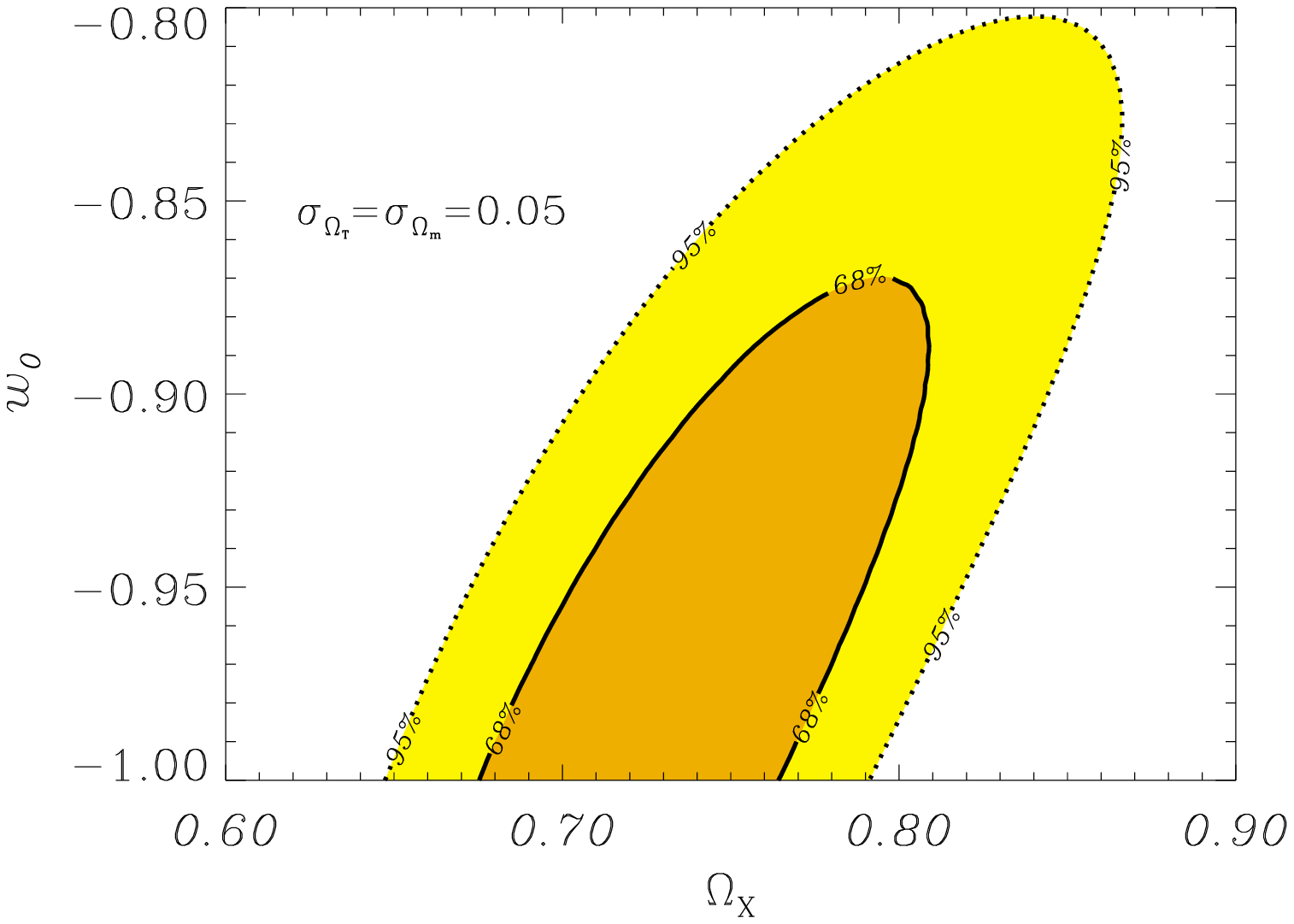}
  \caption{68.3 \% and 95 \% confidence regions for
    $(\Omega_X,w_0)$ in the one-year SNAP scenario assuming prior knowledge
    with $\Omega_m$ and 
    $\Omega_{\rm tot}$ independently Gaussian around their true values and
    $\sigma_{\Omega_m{\rm-prior}}=0.05=\sigma_{\Omega_{\rm tot}{\rm-prior}}$.
    The first figure assumes exact knowledge of ${\cal M}$, while the second
    assumes no prior knowledge of ${\cal M}$.}\label{fig:snap05-06} 
\end{figure}

In figure \ref{fig:snap05-06}, all three parameters
$(\Omega_m,\Omega_X,w_0)$ are allowed to vary, while both $\Omega_m$
and $\Omega_{\rm tot}$ are independently subject to Gaussian priors
with $\sigma_{\Omega_m{\rm-prior}}=0.05$ and
$\sigma_{\Omega_{\rm tot}{\rm-prior}}=0.05$. Comparing the case with
exact ${\cal M}$ to the situation with no prior knowledge, it can be
noted that the uncertainty of the latter mainly grows in $w_0$.

\subsection{Confidence regions for $(w_0,w_1)$}\label{sec:conf-w0-w1}
\label{sec:maor} 

Recently, Maor et al. (\cite{maor}; MBS) considered the problem of
determining the equation of state of the dark energy using supernova 
measurements. In particular,
they investigated an idealised experiment with thousands of supernovae
in the redshift range $z\in[0,2]$, divided into 50 bins. The relative
precision of the luminosity-distance $d_L$ was taken to be 0.6 \% per
bin, which corresponds to a magnitude precision
$\sigma_i=0.006\times5/\ln(10)\approx0.0130$ for each bin.
The equation of state is taken to be linear, $w(z)=w_0+w_1\,z$.
Confidence regions for $(w_0,w_1)$ were determined, using the 
cosmology $\vec{\theta}_{\rm true}=(0.3,0.7,-0.7, 0)$. The log-likelihood
was determined both for an exact $\Omega_m$, and with $\Omega_m$
integrated over $\Omega_{m,{\rm true}}\pm0.1$.

\begin{figure}
  \centering
  \includegraphics[width=\hsize]{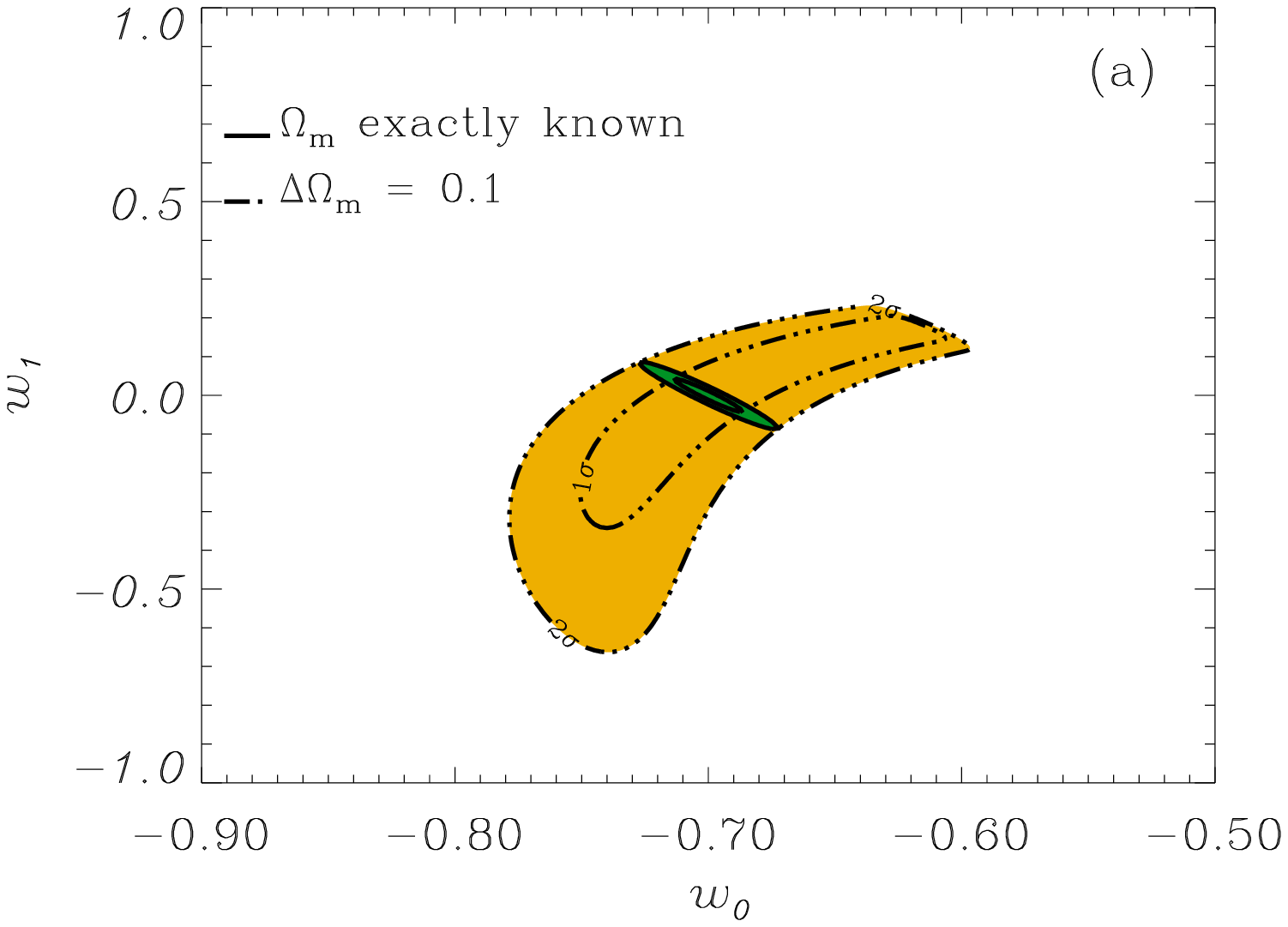} \\
  \includegraphics[width=\hsize]{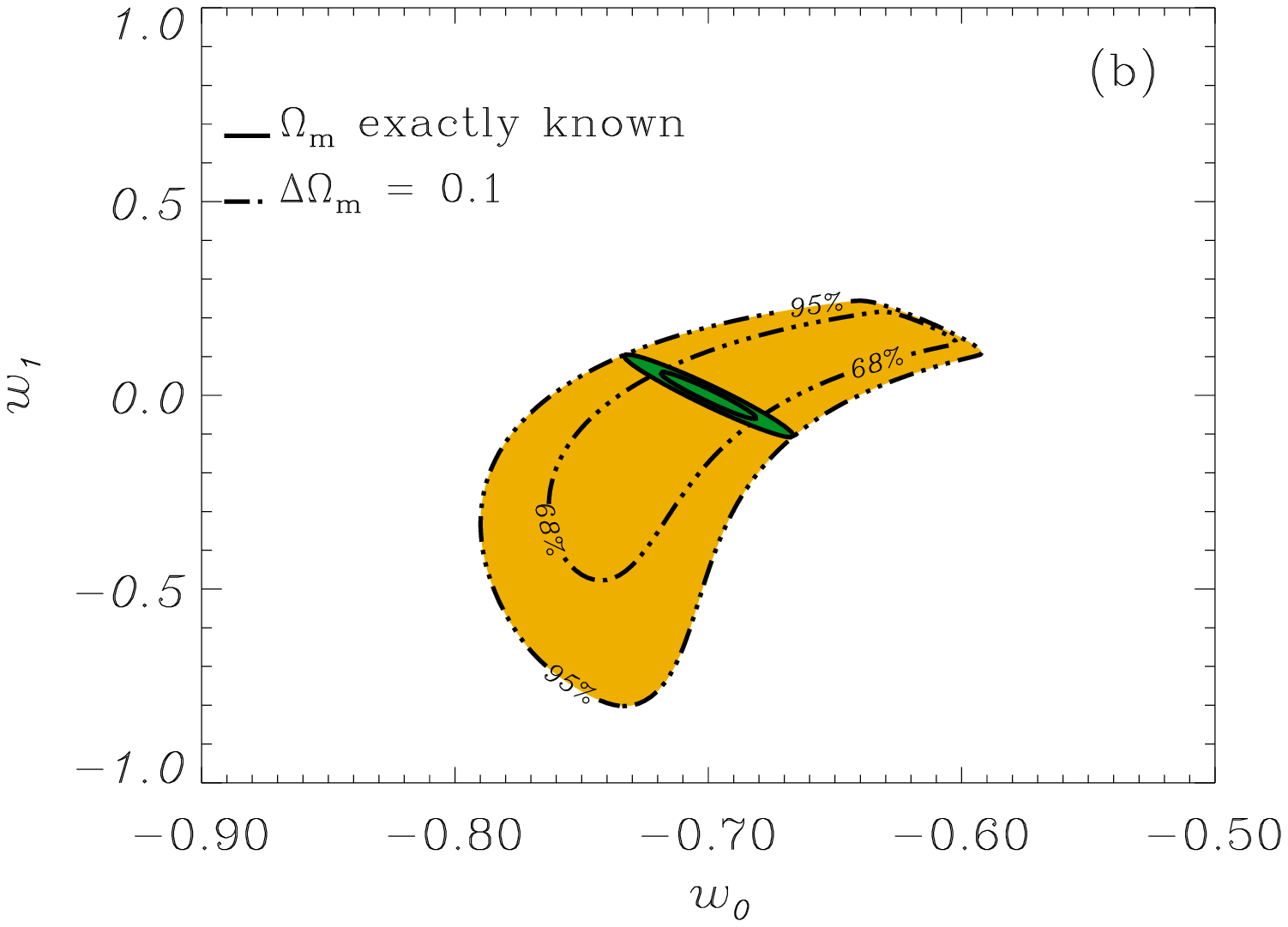}
  \caption{(a) 1$\sigma$ and 2$\sigma$ confidence regions for
    $(w_0,w_1)$ using the scenario of Maor et
    al. (\protect\cite{maor}). The elongated 
    ellipses correspond to the assumption of 
    exact knowledge of $\Omega_m$, while the larger, non-elliptic
    regions assume the prior knowledge that $\Omega_m$ is confined to
    the interval $\Omega_{m,{\rm true}}\pm0.1$. Exact
    knowledge of ${\cal M}$ is assumed. (b) 68.3 \% and 95 \% 
    confidence regions for the same cosmology.}
  \label{fig:maor} 
\end{figure}

Figure \ref{fig:maor} shows our calculation of the
confidence regions for this scenario. This figure should be compared 
with figure 2 of MBS. (Since MBS present one- and two-sigma 
contours, rather than the 68.3 \% and 95 \% confidence regions, we have 
included both cases to facilitate comparison.) There is a 
considerable discrepancy between these figures and MBS\footnote{It has
  come to our attention that MBS used $\sigma_i=0.03$ mag., which
  corresponds to a relative precision in $d_L$ of about 1.4
  \%, and that their contours really correspond to 68.3 \% and 95 \%
  confidence regions (Brustein, private communication). This fully accounts
  for the discrepacy between figures.}.  

\begin{figure}
  \centering
  \includegraphics[width=\hsize]{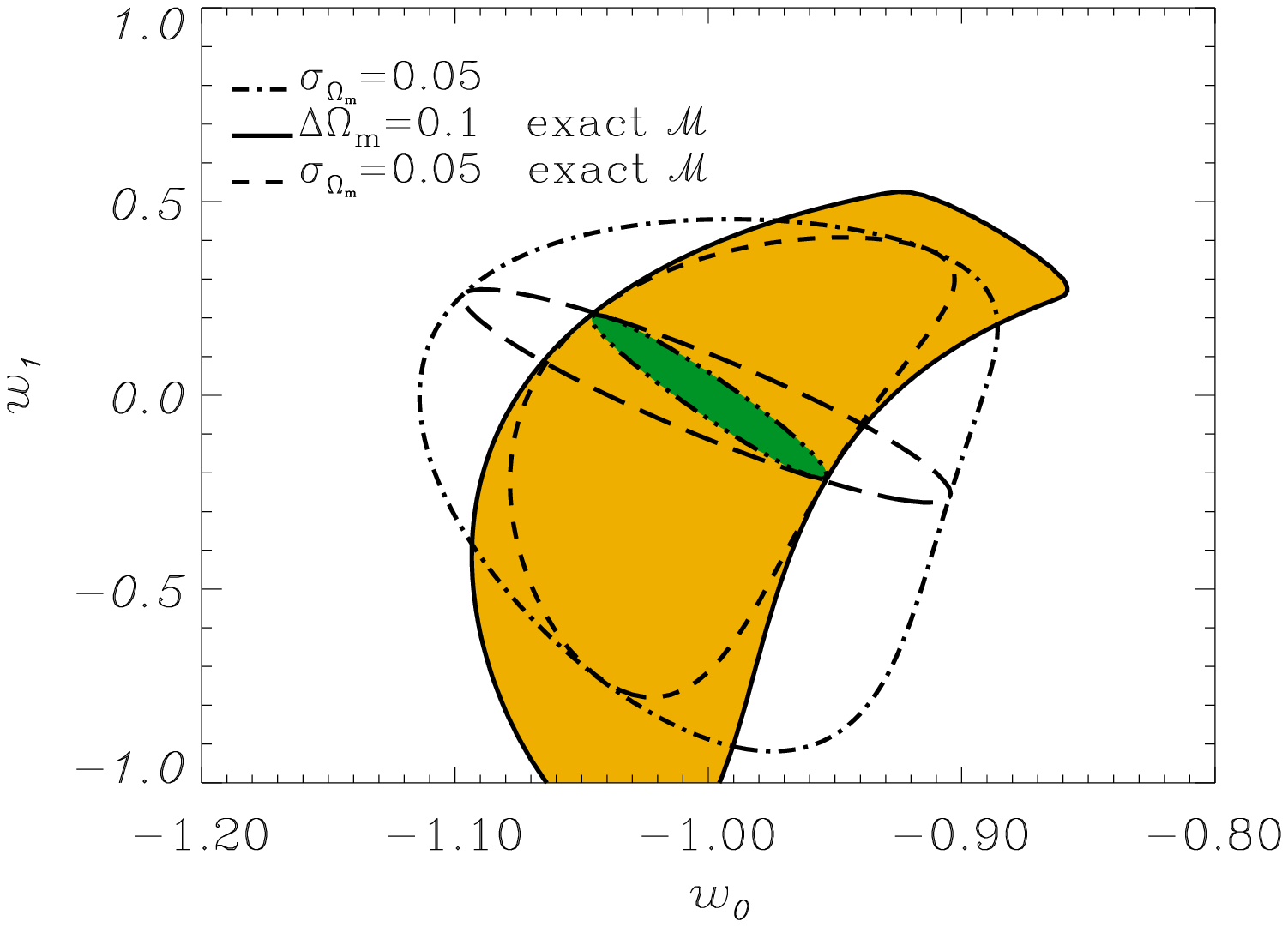}
  \caption{68.3 \% confidence regions for
    $(w_0,w_1)$ in the one-year SNAP scenario. 
    The elongated ellipses correspond to the assumption of 
    exact knowledge of $\Omega_m$: 
    the dash-dot-dot-dotted line is with exact ${\cal M}$ and 
    the long-dashed line corresponds to no knowledge of ${\cal M}$. 
    The larger, non-elliptic regions assume prior knowledge of $\Omega_m$: 
    the dash-dotted line assumes that $\Omega_m$ is known with a Gaussian prior 
    for which $\sigma_{\Omega_m{\rm-prior}}=0.05$; 
    the short-dashed line assumes the same prior and exact
    knowledge of ${\cal M}$;
    finally, the solid line is with $\Omega_m$ confined to the 
    interval $\Omega_{m}\pm0.1$ and exact
    knowledge of ${\cal M}$.}
  \label{fig:snap07-10} 
\end{figure}

\begin{figure}
  \centering
  \includegraphics[width=\hsize]{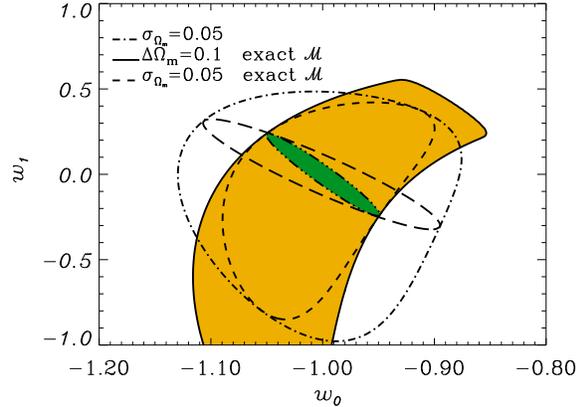}
  \caption{68.3 \%  confidence regions for
    $(w_0,w_1)$ in the one-year SNAP scenario without the 100 events
    for which $z\in[1.2,1.7]$. 
    The elongated ellipses correspond to the assumption of 
    exact knowledge of $\Omega_m$: 
    the dash-dot-dot-dotted line is with exact ${\cal M}$ and 
    the long-dashed line corresponds to no knowledge of ${\cal M}$. 
    The larger, non-elliptic regions assume prior knowledge of $\Omega_m$: 
    the dash-dotted line assumes that $\Omega_m$ is known with a Gaussian prior 
    for which $\sigma_{\Omega_m{\rm-prior}}=0.05$; 
    the short-dashed line assumes the same prior and exact
    knowledge of ${\cal M}$;
    finally, the solid line is with $\Omega_m$ confined to the 
    interval $\Omega_{m}\pm0.1$ and exact
    knowledge of ${\cal M}$.}
  \label{fig:snap20-23} 
\end{figure}

\begin{figure}
  \centering
  \includegraphics[width=\hsize]{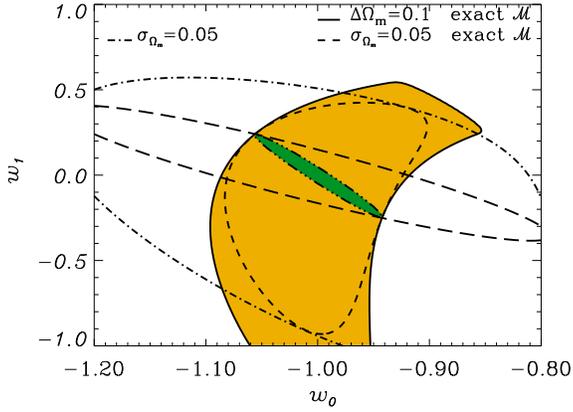}
  \caption{68.3 \% confidence regions for
    $(w_0,w_1)$ in the one-year SNAP scenario with a constant rate
    per co-moving volume for $z\in[0,1.2]$, and the 100 $z\in[1.2,1.7]$ events
    uniformly distributed. 
    The elongated ellipses correspond to the assumption of 
    exact knowledge of $\Omega_m$: 
    the dash-dot-dot-dotted line is with exact ${\cal M}$ and 
    the long-dashed line corresponds to no knowledge of ${\cal M}$. 
    The larger, non-elliptic regions assume prior knowledge of $\Omega_m$: 
    the dash-dotted line assumes that $\Omega_m$ is known with a Gaussian prior 
    for which $\sigma_{\Omega_m{\rm-prior}}=0.05$; 
    the short-dashed line assumes the same prior and exact
    knowledge of ${\cal M}$;
    finally, the solid line is with $\Omega_m$ confined to the 
    interval $\Omega_{m}\pm0.1$ and exact
    knowledge of ${\cal M}$.}
  \label{fig:snap25-28} 
\end{figure}

In conclusion, it seems to us that this scenario enables
a better constraining of $(w_0,w_1)$ than was previously
anticipated by MBS. However, the scenario assumes that more than 6000
supernovae uniformly distributed over a rather optimistic redshift
range are observed. Consequently, in 
this section, we calculate $(w_0,w_1)$ confidence regions for
the cosmology of MBS, using the weaker precision and a smaller
redshift range assumed in the SNAP proposal (\cite{snap}),
see figure \ref{fig:snap11-14} below. However, we focus our attention on
the fiducial cosmology of SNAP:  
$\vec{\theta}_{\rm true}=(0.28,0.72,-1, 0)$ 
(see figures \ref{fig:snap07-10} -- \ref{fig:snap25-28} and 
\ref{fig:snap15-18}).

We consider the ability to determine the equation of state of
the dark energy to linear order, $w(z)=w_0+w_1\,z$. We will assume
flatness, $\Omega_{\rm tot}=1$, and impose some prior knowledge of
$\Omega_m$. Figure \ref{fig:snap07-10} shows confidence regions for
various assumptions regarding ${\cal M}$ and $\Omega_m$. We mainly
consider a Gaussian prior with $\sigma_{\Omega_m{\rm-prior}}=0.05$.
The uniform prior with $\Omega_m\in\Omega_{m,{\rm true}}\pm0.1$ is
considered in the case of exact knowledge of ${\cal M}$,
since this is the situation considered by MBS.
In figure \ref{fig:snap20-23}, the few high-$z$ supernovae have been
excluded. When ${\cal M}$ is exactly known, these are not so important
in determining $(w_0,w_1)$ as 
they are for $(\Omega_m,\Omega_X)$, basically because $\Omega_X$
becomes less significant with increasing redshift. However, note that
the high-$z$ events make some difference when ${\cal M}$ is poorly known. 
Figure \ref{fig:snap25-28} shows the situation when the supernovae at
$z\in[0,1.2]$ are distributed according to a constant rate per
co-moving volume element. As can be expected, uncertainties are not affected
when ${\cal M}$ is considered to be exactly known, but degrade
considerably with no prior information of ${\cal M}$.
Figure \ref{fig:snap11-14} considers the same scenario as in figure
\ref{fig:snap07-10} as regards
precision and priors for $\Omega_m$, but uses the fiducial cosmology
of MBS.

With the priors for $\Omega_m$ assumed in figures \ref{fig:snap07-10}
-- \ref{fig:snap11-14}, the equation-of-state parameters are rather
poorly constrained by one year of SNAP data, especially when
${\cal M}$ is left unspecified. In order to see what SNAP can achieve
over its expected three years of operation, we calculate the
confidence regions for thrice as many supernovae. Priors for
$\Omega_m$ are Gaussian with $\sigma_{\Omega_m{\rm-prior}}=0.05$ as
before, and we also consider $\sigma_{\Omega_m{\rm-prior}}=0.015$. The
latter is consistent with the estimated precision of a hypothetical
ground-based $10^\circ\times10^\circ$ weak-lensing survey
(van Waerbeke et al. \cite{waerbeke}). (As discussed in this reference there 
is a weak dependence 
of $\Omega_X$ in these estimates of $\Omega_m$. We will not pursue this 
further here.) Uncertainties when $\Omega_m$ is exactly known
(elliptic contours) improve the expected factor $1/\sqrt{3}$ as
compared with the one-year scenario (compare with figure
\ref{fig:snap07-10}).
For an $\Omega_m$ prior with
$\sigma_{\Omega_m{\rm-prior}}=0.05$, confidence regions still span
considerable parts of the parameter space.
However, with the sharper prior, uncertainties in $w_0$ and
$w_1$ go down to $w_0=-1\pm0.02$, $w_1=0^{+0.13}_{-0.15}$ with an exact
${\cal M}$, and are still reasonable when imposing no prior knowledge
of ${\cal M}$: $w_0=-1\pm0.04$, $w_1=0^{+0.15}_{-0.17}$. Thus, it
seems to us that three years of SNAP data backed up with independent
high-precision observations of $\Omega_m$ can constrain the nature of
the dark energy quite well.
Note that in the above calculations we implicitly assume that the
universe is known to be flat with high accuracy, since we
have imposed $\Omega_{\rm tot}=1$.

Next, we turn our attention to the impact of different
redshift distributions of supernovae on the confidence region
in the  ($w_0,w_1$) parameter space. The interval $z\in[0,2]$ is
divided into 8 subsets containing 500 supernovae each uniformly distributed 
in redshift as:
$z_1=[0,0.25]$, $z_2=[0.25,0.5]$, $z_3=[0.5,0.75]$,.., $z_8=[1.75,2.0]$. 
We then compose 4 different experimental situations
where in each case 2000 supernovae are measured ($\Delta m$ = 0.16 mag /SN) 
sampling events from $[z_1,z_2,z_3,z_4]$,  $[z_1,z_2,z_7,z_8]$, 
$[z_2,z_4,z_6,z_8]$ and  $[z_5,z_6,z_7,z_8]$, as shown in figure \ref{fig:zdep}
for the case $\vec{\theta}=(0.3,0.7,-1,0)$ where the mass
energy density is given a uniform prior with 
$\Delta\Omega_m=0.1$.
Clearly, a wide range of supernova redshifts is more advantageous than 
only data above or below $z=1$. 

\begin{figure}
  \centering
  \includegraphics[width=\hsize]{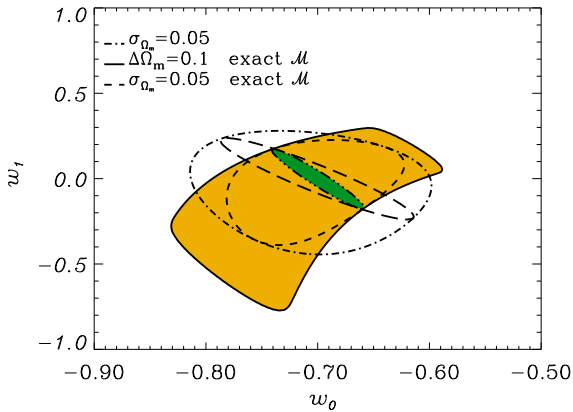}
  \caption{68.3 \% confidence regions for
    $(w_0,w_1)$ assuming the precision of the one-year SNAP scenario,
    but the cosmology of Maor et al. (\protect\cite{maor}). 
    The elongated ellipses correspond to the assumption of 
    exact knowledge of $\Omega_m$: 
    the dash-dot-dot-dotted line is with exact ${\cal M}$ and 
    the long-dashed line corresponds to no knowledge of ${\cal M}$. 
    The larger, non-elliptic regions assume prior knowledge of $\Omega_m$: 
    the dash-dotted line assumes that $\Omega_m$ is known with a Gaussian prior 
    for which $\sigma_{\Omega_m{\rm-prior}}=0.05$; 
    the short-dashed line assumes the same prior and exact
    knowledge of ${\cal M}$;
    finally, the solid line is with $\Omega_m$ confined to the 
    interval $\Omega_{m}\pm0.1$ and exact
    knowledge of ${\cal M}$.}
  \label{fig:snap11-14} 
\end{figure}

\begin{figure}
  \centering
  \includegraphics[width=\hsize]{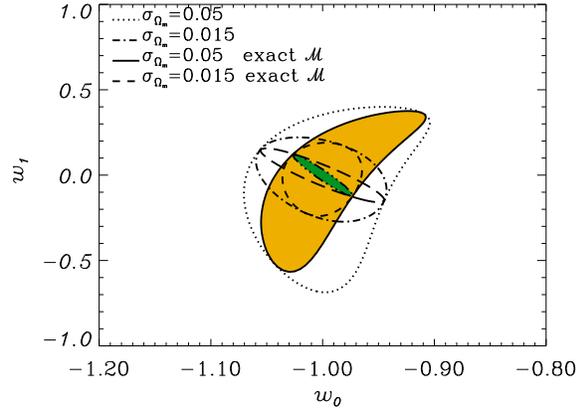}
  \caption{68.3 \% confidence regions for
    $(w_0,w_1)$ in the three-year SNAP scenario. 
    The elongated ellipses correspond to the assumption of 
    exact knowledge of $\Omega_m$: 
    the dash-dot-dot-dotted line  is with exact ${\cal M}$ and 
    the long-dashed line corresponds to no knowledge of ${\cal M}$. 
    The larger, non-elliptic regions assume 
    Gaussian prior knowledge of $\Omega_m$: 
    the dotted line is with $\sigma_{{\Omega_m}{\rm-prior}}=0.05$, 
    while the dash-dotted line is with $\sigma_{\Omega_m{\rm-prior}}=0.015$. 
    The solid and short-dashed lines assume exact knowledge of ${\cal M}$
    with the same $\Omega_m$ priors as above.}
  \label{fig:snap15-18}  
\end{figure}

\begin{figure}
  \centering
  \includegraphics[width=\hsize]{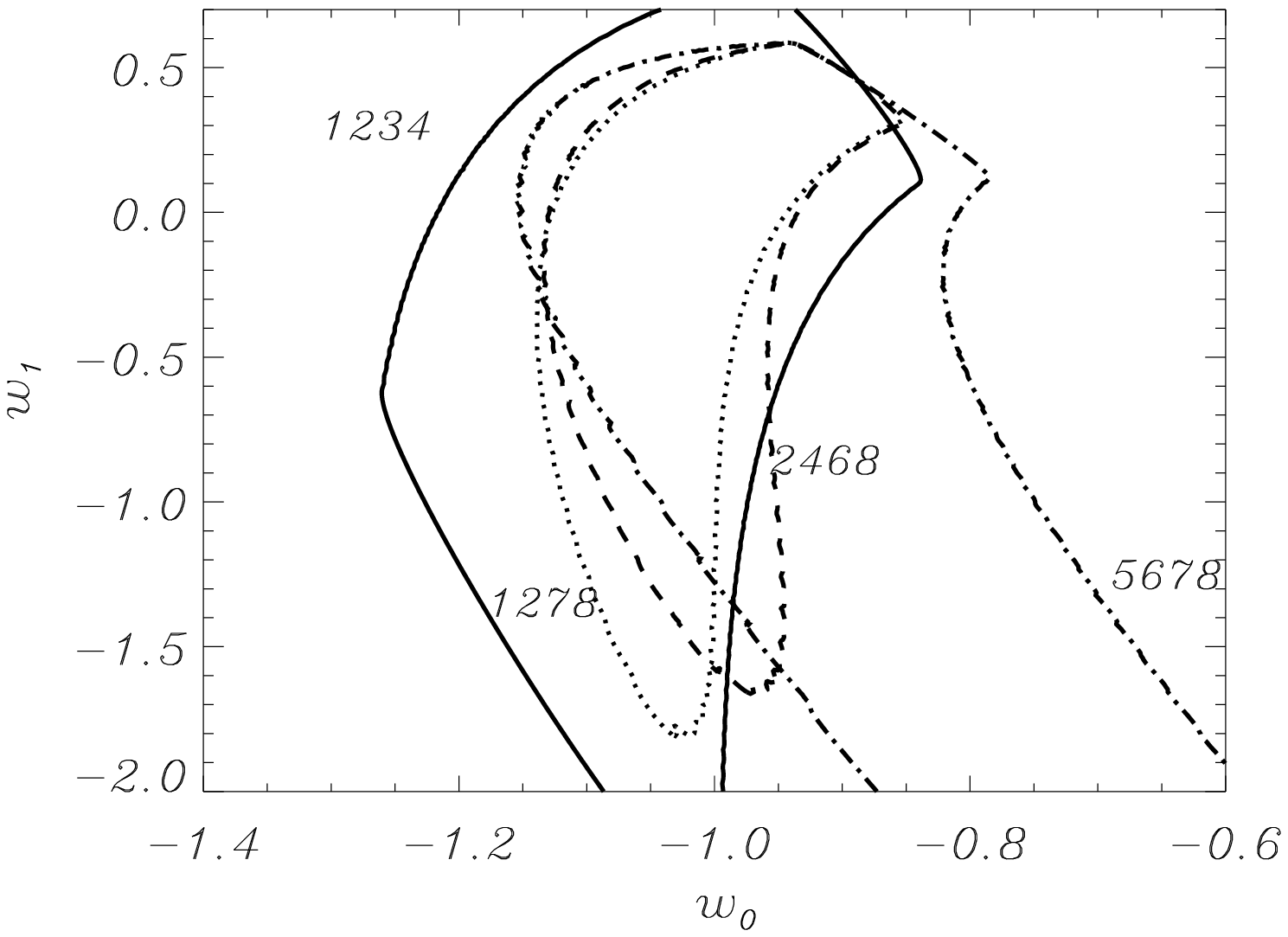}
  \caption{68.3 \% confidence regions for
    $(w_0,w_1)$ with 2000 supernovae. 
    The importance of wide redshift coverage is demonstrated by 
    simulating four different synthetic experiments, all consisting of 
    2000 SNe: 1234 includes SNe uniformly distributed in $z\in[0,1]$, 
    experiment 5678 has only SNe uniformly distributed in $z\in[1,2]$, 
    experiment 1278 has supernovae in uniformly distributed in two bins,
    $z\in[0,0.5]$ and $z\in[1.5,2]$. Finally, experiment 2468 includes 
    four bins: $z\in[0.25,0.5]$, $z\in[0.75,1]$, $z\in[1.25,1.5]$ and 
    $z\in[1.75,2]$. Clearly, the two experiments with the widest redshift 
    coverage provide the best constraints.}
  \label{fig:zdep}  
\end{figure}

\section{Effect of adding a small sample of supernovae}\label{sec:add}

To further illustrate the importance of a small number of high-redshift
events, we have performed Fisher analyses 
(see appendix \ref{app:fisher} and (Astier \cite{astier})) 
to investigate the effect of adding 100 supernovae to a
large initial sample at lower redshift. We do this for initially 2000
supernovae uniformly distributed at $z\in[0,1.2]$. To emphasize the
importance of events at very low redshift, we do the same exercise for
initially 2000 supernovae uniformly distributed at $z\in[0.2,1.2]$.
Since the effects depend significantly on the underlying cosmological
model, we investigate three models: the fiducial model of the SNAP
proposal (\cite{snap}): $\vec{\theta}_{\rm true}=(0.28,0.72,-1,0)$,
a quintessence model derived from supergravity considerations 
(Brax \& Martin \cite{brax})
$\vec{\theta}_{\rm true}=(0.28,0.72,-0.8,0.3)$, and the model used by
Maor et al. (\cite{maor}) $\vec{\theta}_{\rm true}=(0.3,0.7,-0.7,0)$.

\begin{figure}
  \centering
  \includegraphics[width=\hsize]{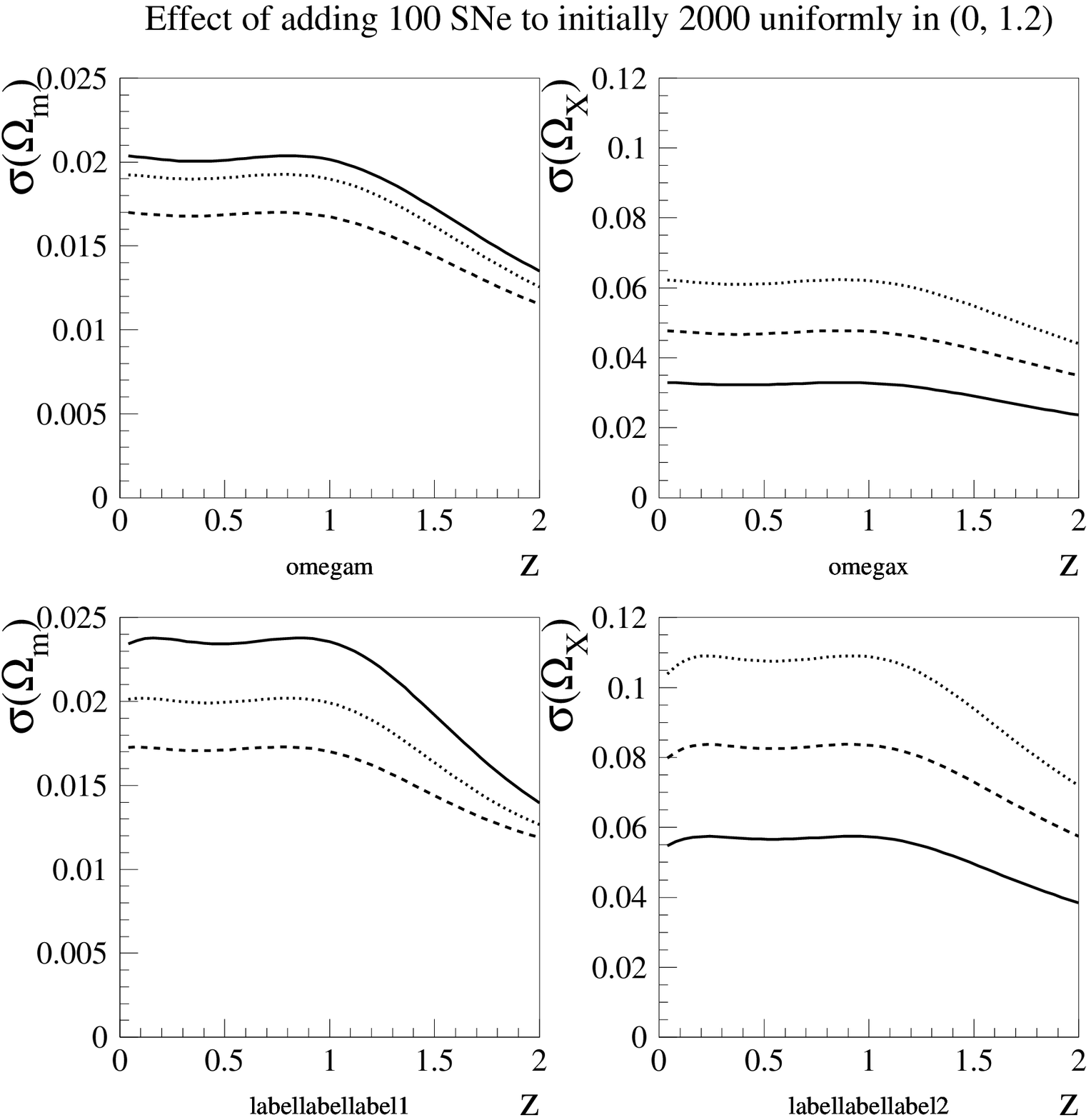}
  \psfrag{label1}{$\Omega_m$, $\cal{M}$ unknown}
  \caption{The effect on $\sigma_{\Omega_m}$ and 
    $\sigma_{\Omega_X}$
    when 100 supernovae are added at a specific redshift $z\in[0,2]$.
    The original sample consists of 2000 supernovae uniformly
    distributed over $z\in[0,1.2]$. Solid lines correspond to the SNAP
    fiducial model $(\Omega_m,\Omega_X,w_0,w_1)=(0.28,0.72,-1,0)$,
    dashed lines correspond to
    $(\Omega_m,\Omega_X,w_0,w_1)=(0.28,0.72,-0.8,0.3)$,
    and dotted lines correspond to
    $(\Omega_m,\Omega_X,w_0,w_1)=(0.3,0.7,-0.7,0)$.}\label{fig:add01-06}  
\end{figure}

\begin{figure}
  \centering
  \includegraphics[width=\hsize]{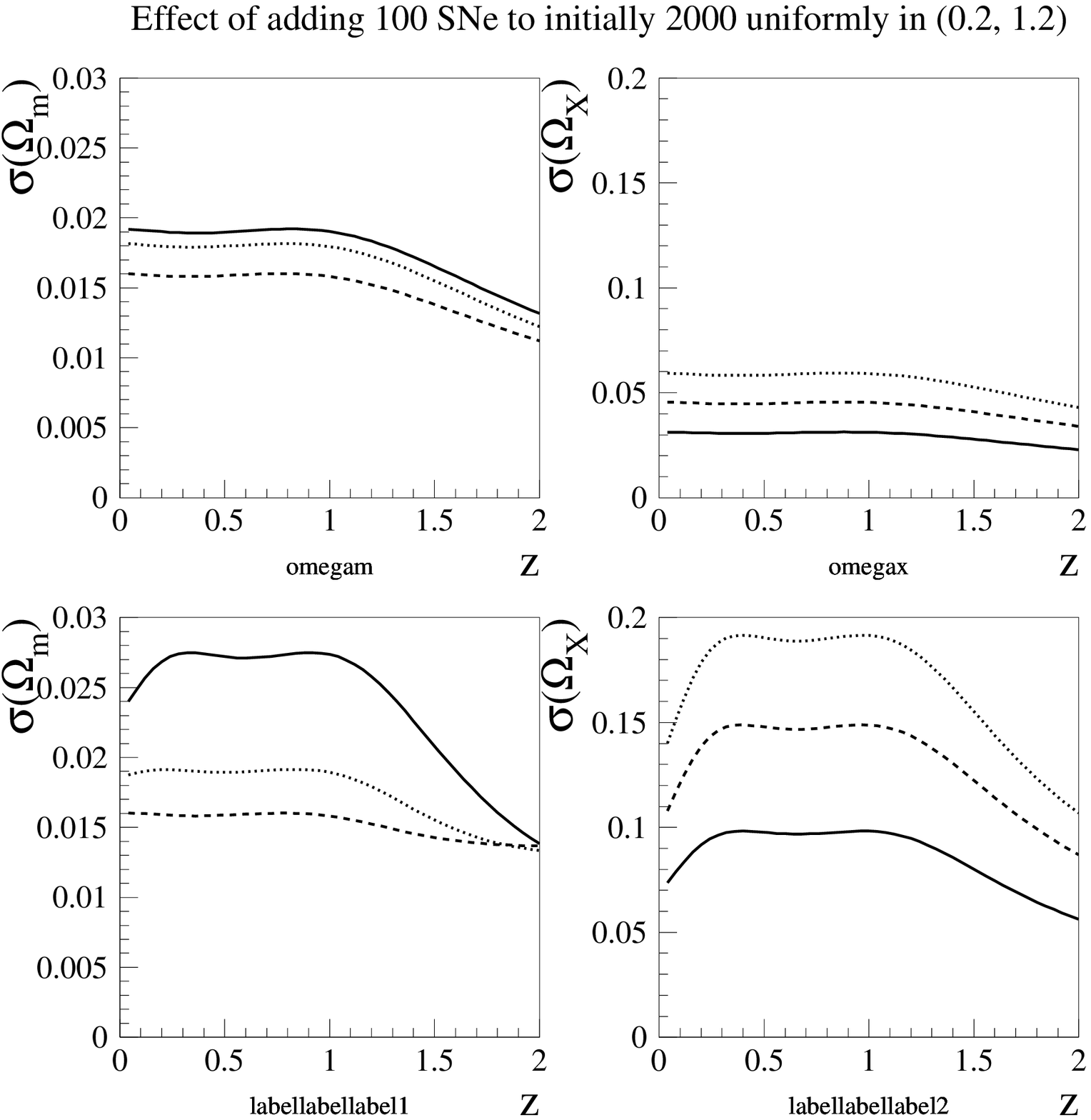}
  \caption{The effect on $\sigma_{\Omega_m}$ and 
    $\sigma_{\Omega_X}$
    when 100 supernovae are added at a specific redshift $z\in[0,2]$.
    The original sample consists of 2000 supernovae uniformly
    distributed over $z\in[0.2,1.2]$. Solid lines correspond to the SNAP
    fiducial model $(\Omega_m,\Omega_X,w_0,w_1)=(0.28,0.72,-1,0)$,
    dashed lines correspond to
    $(\Omega_m,\Omega_X,w_0,w_1)=(0.28,0.72,-0.8,0.3)$,
    and dotted lines correspond to
    $(\Omega_m,\Omega_X,w_0,w_1)=(0.3,0.7,-0.7,0)$.}\label{fig:add07-12}  
\end{figure}

Figures \ref{fig:add01-06} and \ref{fig:add07-12} show the effect on
the errors of $\Omega_m$ and $\Omega_X$ when adding 100 supernovae to
the samples outlined above. As expected, high redshifts pay off when
determining $\Omega_m$ and $\Omega_X$, but in case the knowledge
of ${\cal M}$ is poor, it is also important to fill the low-redshift region.
Note that the curves for exact ${\cal M}$
have two minima ($z_{\rm max}$ and one intermediate redshift), while
those where ${\cal M}$ is unknown have three ($z_{\rm min}$,
$z_{\rm max}$ and one intermediate redshift). This is only a
manifestation of the fact that the optimum redshift distribution with
$n$ parameters consists of $n$ $\delta$ functions (Astier \cite{astier}).
(When priors are imposed this may no longer be the case.) Furthermore, for
each curve there are two values of the redshift where it is totally
ineffectual to add more events.

\begin{figure}
  \centering
  \includegraphics[width=\hsize]{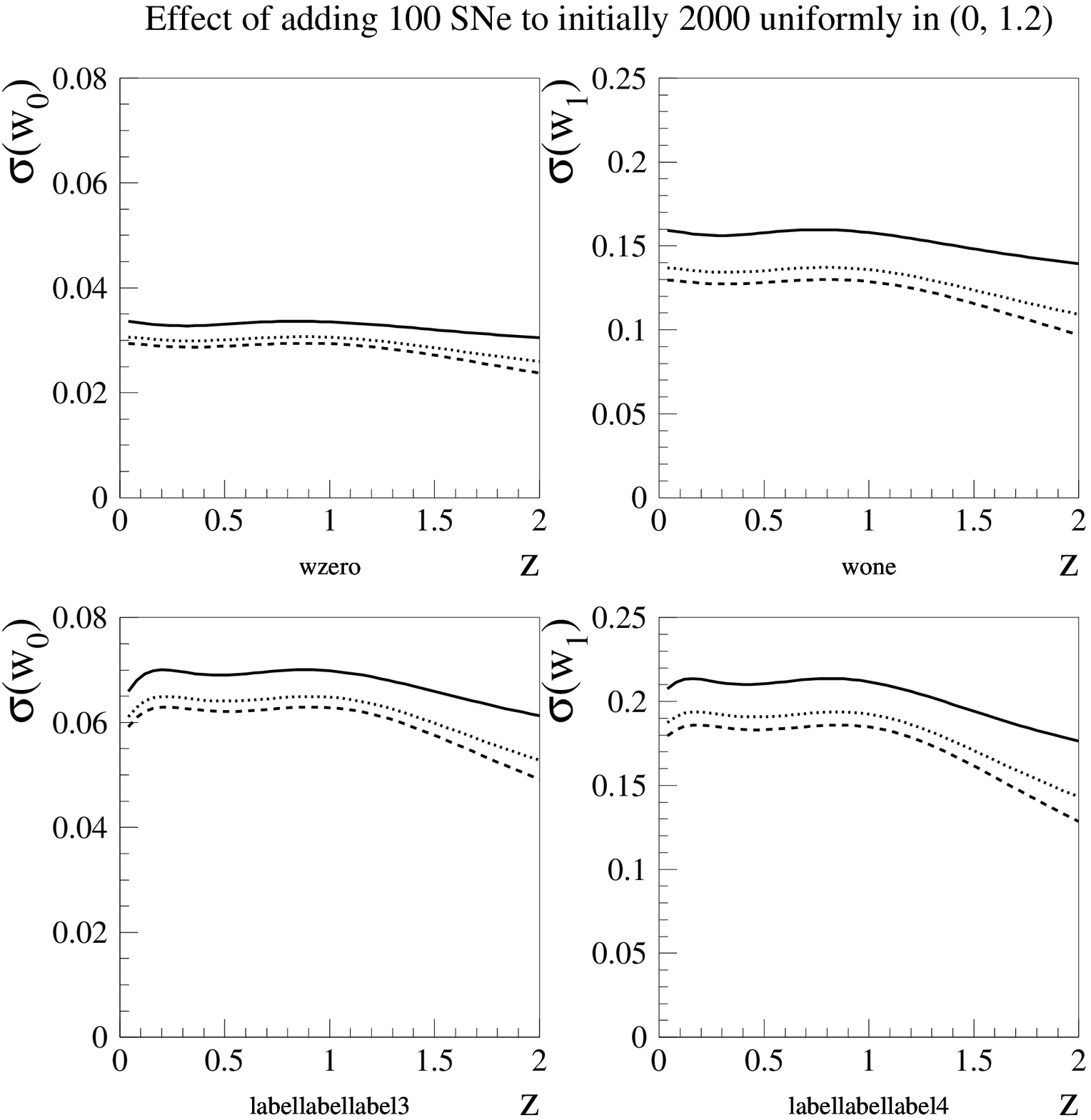}
  \caption{The effect on $\sigma_{w_0}$ and $\sigma_{w_1}$
    when 100 supernovae are added at a specific redshift $z\in[0,2]$.
    The original sample consists of 2000 supernovae uniformly
    distributed over $z\in[0,1.2]$. $\Omega_m$ and $\Omega_X$ are
    assumed to be exactly known. Solid lines correspond to the SNAP
    fiducial model $(\Omega_m,\Omega_X,w_0,w_1)=(0.28,0.72,-1,0)$,
    dashed lines correspond to
    $(\Omega_m,\Omega_X,w_0,w_1)=(0.28,0.72,-0.8,0.3)$,
    and dotted lines correspond to
    $(\Omega_m,\Omega_X,w_0,w_1)=(0.3,0.7,-0.7,0)$.}\label{fig:add13-18}  
\end{figure}

\begin{figure}
  \centering
  \includegraphics[width=\hsize]{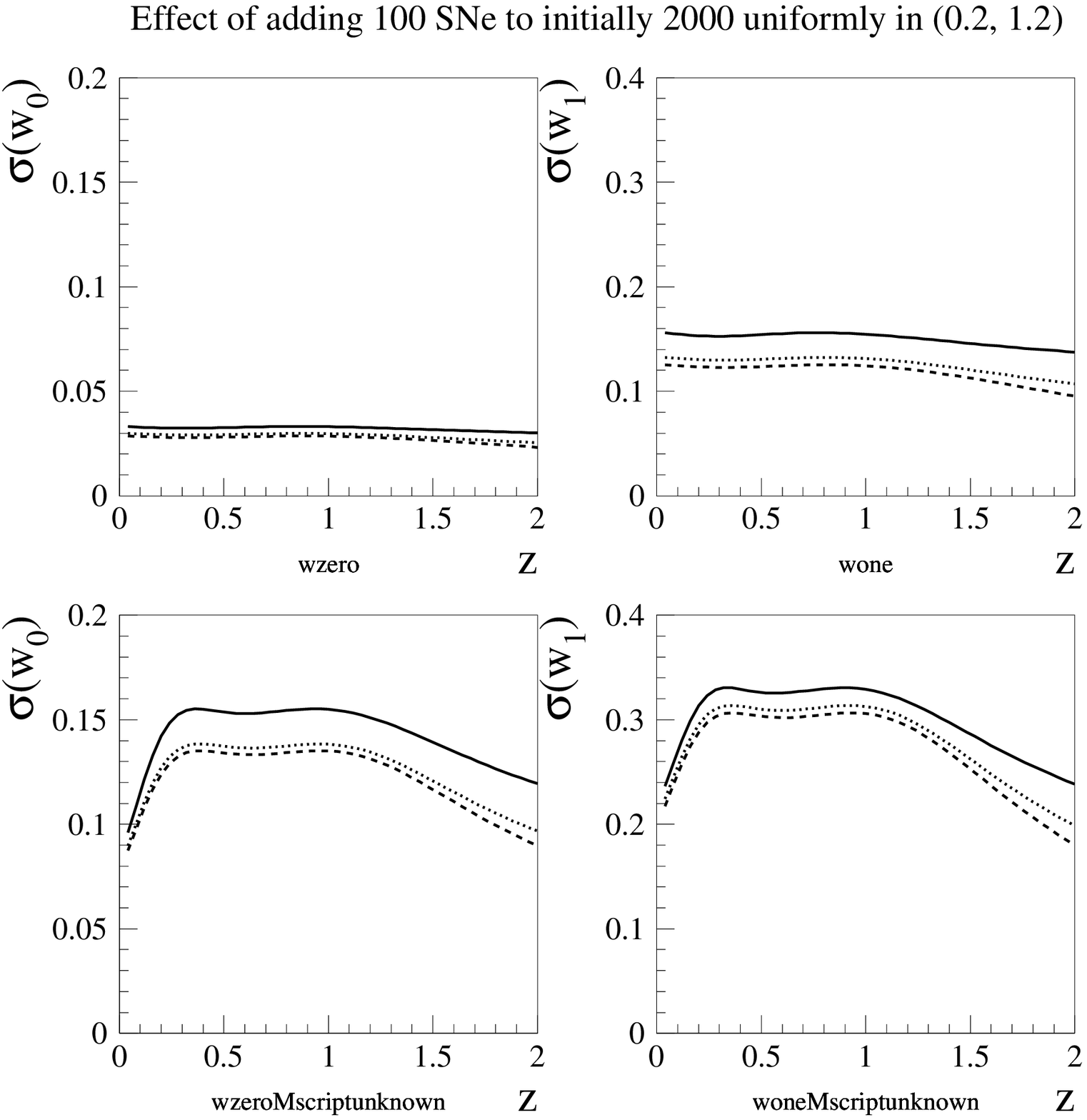}
  \caption{The effect on $\sigma_{w_0}$ and $\sigma_{w_1}$
    when 100 supernovae are added at a specific redshift $z\in[0,2]$.
    The original sample consists of 2000 supernovae uniformly
    distributed over $z\in[0.2,1.2]$. $\Omega_m$ and $\Omega_X$ are
    assumed to be exactly known. Solid lines correspond to the SNAP
    fiducial model $(\Omega_m,\Omega_X,w_0,w_1)=(0.28,0.72,-1,0)$,
    dashed lines correspond to
    $(\Omega_m,\Omega_X,w_0,w_1)=(0.28,0.72,-0.8,0.3)$,
    and dotted lines correspond to
    $(\Omega_m,\Omega_X,w_0,w_1)=(0.3,0.7,-0.7,0)$.}\label{fig:add19-24}  
\end{figure}

\begin{figure}
  \centering
  \includegraphics[width=\hsize]{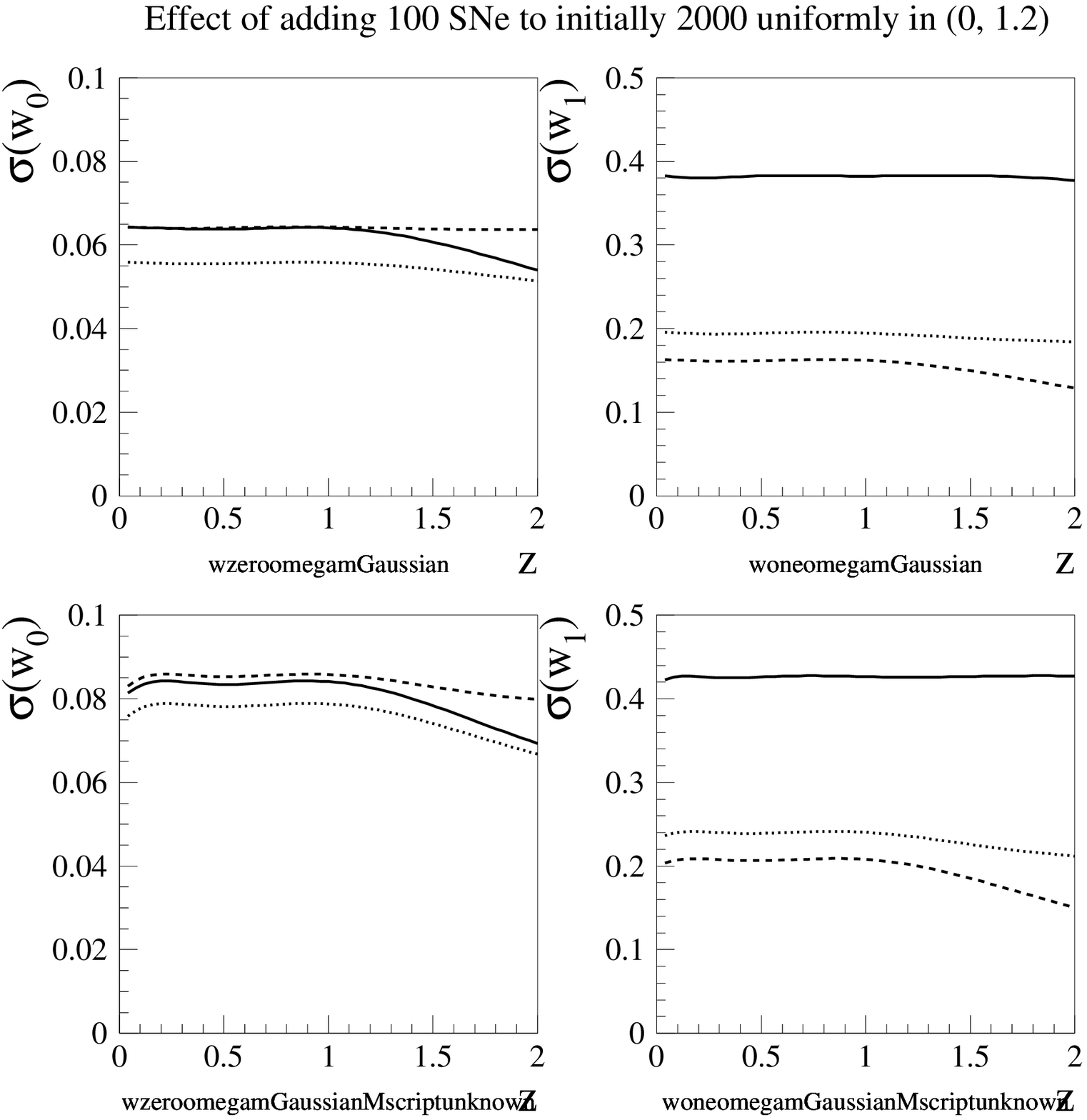}
  \caption{The effect on $\sigma_{w_0}$ and $\sigma_{w_1}$
    when 100 supernovae are added at a specific redshift $z\in[0,2]$.
    The original sample consists of 2000 supernovae uniformly
    distributed over $z\in[0,1.2]$. $\Omega_X$ is
    assumed to be exactly known, while $\Omega_m$ is known within
    $\sigma_{\Omega_m{\rm-prior}}=0.05$.
    Solid lines correspond to the SNAP
    fiducial model $(\Omega_m,\Omega_X,w_0,w_1)=(0.28,0.72,-1,0)$,
    dashed lines correspond to
    $(\Omega_m,\Omega_X,w_0,w_1)=(0.28,0.72,-0.8,0.3)$,
    and dotted lines correspond to
    $(\Omega_m,\Omega_X,w_0,w_1)=(0.3,0.7,-0.7,0)$.}\label{fig:add25-30}  
\end{figure}

\begin{figure}
  \centering
  \includegraphics[width=\hsize]{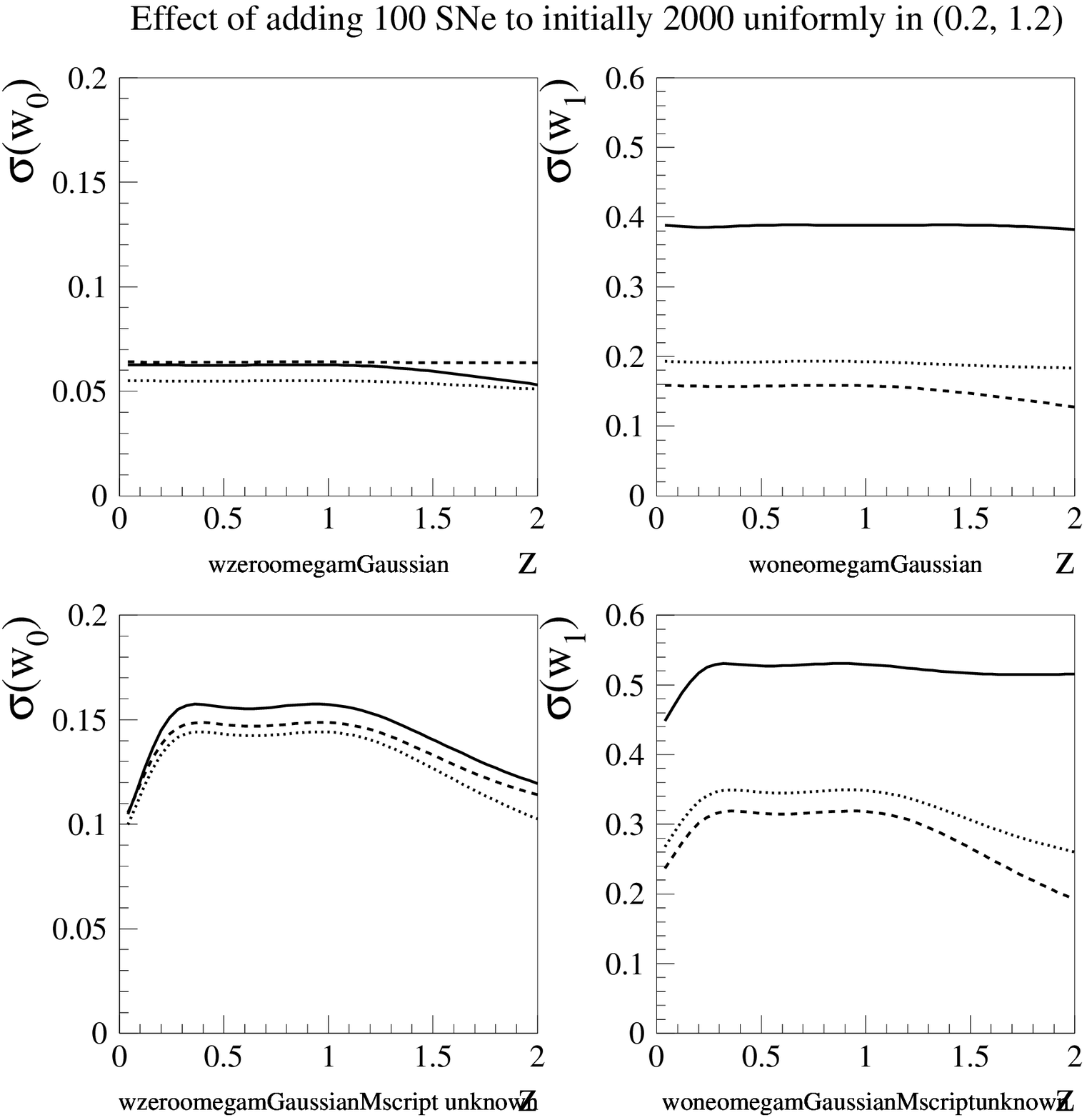}
  \caption{Effect on $\sigma_{w_0}$ and $\sigma_{w_1}$
    when 100 supernovae are added at a specific redshift $z\in[0,2]$.
    The original sample consists of 2000 supernovae uniformly
    distributed over $z\in[0.2,1.2]$. $\Omega_X$ is
    assumed to be exactly known, while $\Omega_m$ is known within
    $\sigma_{\Omega_m{\rm-prior}}=0.05$.
    Solid lines correspond to the SNAP
    fiducial model $(\Omega_m,\Omega_X,w_0,w_1)=(0.28,0.72,-1,0)$,
    dashed lines correspond to
    $(\Omega_m,\Omega_X,w_0,w_1)=(0.28,0.72,-0.8,0.3)$,
    and dotted lines correspond to
    $(\Omega_m,\Omega_X,w_0,w_1)=(0.3,0.7,-0.7,0)$.}\label{fig:add31-36}  
\end{figure}

Figures \ref{fig:add13-18} -- \ref{fig:add31-36} assume a flat
universe, and consider $(w_0,w_1)$ for the same initial
distributions. In figures \ref{fig:add13-18} and \ref{fig:add19-24}
$\Omega_m$ is exactly known, while in figures \ref{fig:add25-30} and
\ref{fig:add31-36} a Gaussian prior with $\sigma_{\Omega_m{\rm-prior}}=0.05$ 
is imposed. The pay-off with high-redshift events is not as great as when
determining $(\Omega_m,\Omega_X)$. In particular, note that the
cosmological-constant model is the worst case of the scenarios we
have considered. 

\psfrag{labellabellabel1}{\hspace{-1em}\tiny{$\cal{M}$ unknown}}
\psfrag{labellabellabel2}{\hspace{-1em}\tiny{$\cal{M}$ unknown}}
\psfrag{labellabellabel3}{\tiny{$\cal{M}$ unknown}}
\psfrag{labellabellabel4}{\tiny{$\cal{M}$ unknown}}
\psfrag{wzeroMscriptunknown}{\tiny{$\cal{M}$ unknown}}
\psfrag{woneMscriptunknown}{\tiny{$\cal{M}$ unknown}}
\psfrag{wzeroomegamGaussian}{\tiny{$\Omega_m$ Gaussian}}
\psfrag{woneomegamGaussian}{\tiny{$\Omega_m$ Gaussian}}
\psfrag{wzeroomegamGaussianMscript unknown}{\tiny{$\Omega_m$ Gaussian, 
  $\cal{M}$ unk.}}
\psfrag{wzeroomegamGaussianMscriptunknown}{\tiny{$\Omega_m$ Gaussian, 
  $\cal{M}$ unk.}}
\psfrag{woneomegamGaussianMscriptunknown}{\tiny{$\Omega_m$ Gaussian, 
  $\cal{M}$ unk.}}
\psfrag{omegam}{\tiny{}}
\psfrag{omegax}{\tiny{}}
\psfrag{wzero}{\tiny{}}
\psfrag{wone}{\tiny{}}

\section{Lensing bias}\label{sec:lens}

So far, the analysis has not taken into account any systematic errors
in the magnitude measurements. However, there are several possible
mechanisms that can give rise to redshift-dependent systematics:
attenuation by ``gray dust'' in the intergalactic medium would cause
distant sources to look fainter than they really are, and evolutionary
effects $M=M(z)$ of the absolute magnitude of supernovae Ia are
currently not well-known. Furthermore, the effects of gravitational lensing
increase with redshift, and the corresponding magnitude distributions
become markedly non-Gaussian for sources at high redshift. 

\begin{figure}
  \centering
  \includegraphics[width=\hsize]{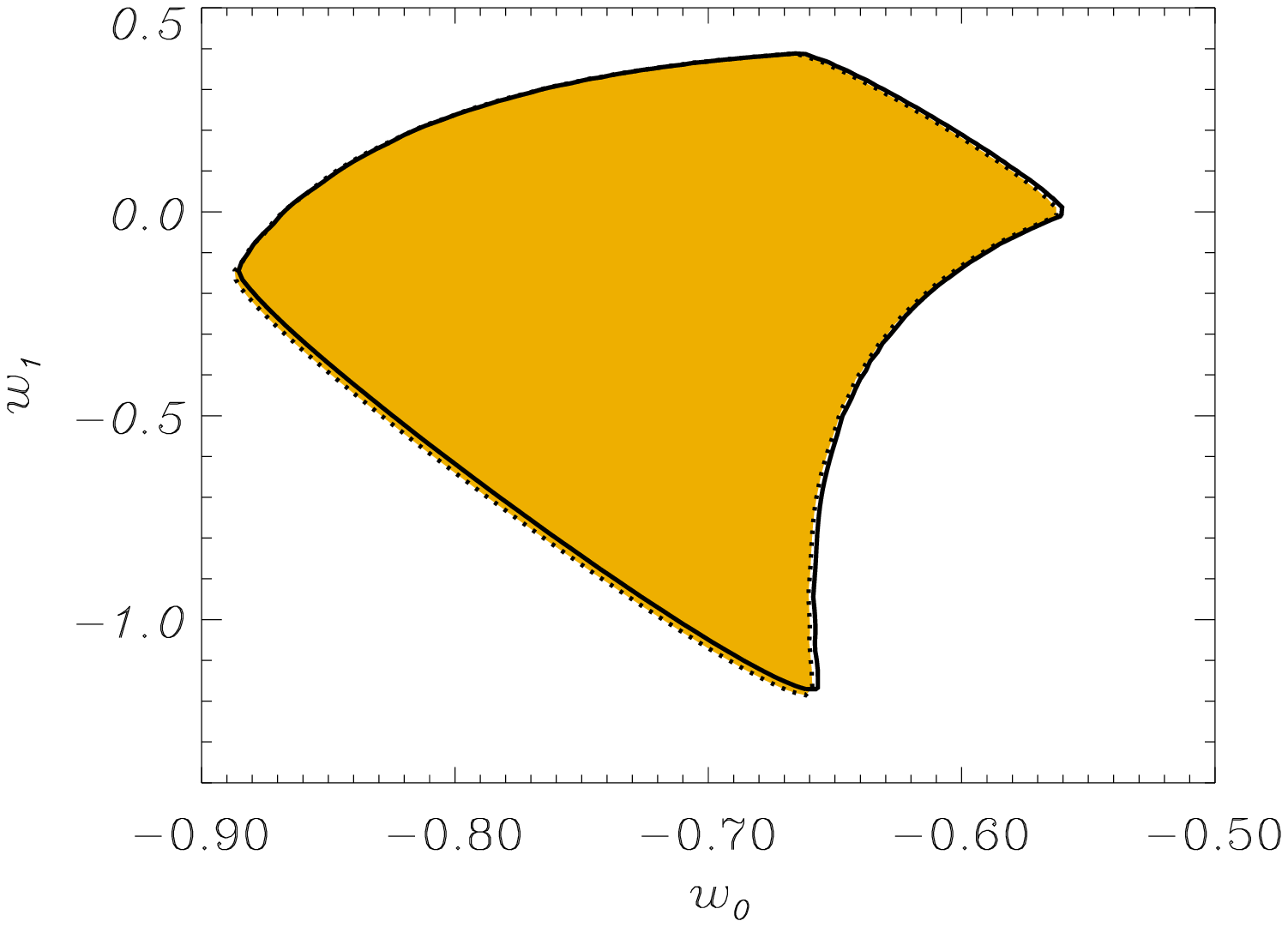}
  \caption{68.3 \% confidence region for
    $(w_0,w_1)$ for one year of SNAP for the cosmology in \cite{maor}, 
central values at $w_0=-0.7$ and $w_1=0$ (solid line). 
The dashed curve incorporates the bias effects form gravitational lensing
magnification.}
  \label{fig:lens}  
\end{figure}

We have investigated the effects from gravitational lensing by using the 
method of Holz \& Wald (\cite{hw}), see further 
(Bergstr\"om et al. \cite{lens}). The 
inhomogeneities are modelled as halos with the density profile as proposed by 
Navarro et al. (\cite{nfw}). 
We consider the cosmology examined by Maor et al. (\cite{maor}),
$\vec{\theta}=(0.3,0.7,-0.7,0)$, and use the redshift distribution
given by table 7.2 in the SNAP proposal (\cite{snap}).
Note that this distribution is different from the ones used previously.
Figure \ref{fig:lens} shows the lensing effects in the $(w_0,w_1)$ space.
In this particular parameter space the lensing effects are negligible 
compared with the intrinsic uncertainty in the $(w_0,w_1)$ measurements. 
However, sizable effects have to be considered for the
$(\Omega_m,\Omega_X)$ parameter space, especially if $\Omega_m$
contains a significant fraction of point-like objects, such as MACHOs
(Amanullah et al. \cite{ramme}). 

\section{Discussion}\label{sec:discuss}

This analysis stresses the importance of combining independent
estimations of the cosmological parameters in order to probe the
nature of the dark energy as accurately as possible. For instance, we
conclude that a mission for observing supernovae over a large redshift
range, such as the SuperNova/Acceleration Probe (SNAP), can give
reasonable constraints on the equation of state of the dark energy,
provided three years of observational data and good prior knowledge of
the geometry and matter density of the universe. To exemplify, we expect 
SNAP to be
able to determine the parameters in a linear equation of state
$w(z)=w_0+w_1\,z$ to within $\pm0.04$ for $w_0$ and
${}^{+0.15}_{-0.17}$ for $w_1$ (one-parameter one-sigma levels),
assuming a flat universe, the matter energy density known with
$\sigma_{\Omega_m{\rm-prior}}\pm0.015$, but no
prior knowledge imposed on the intercept ${\cal M}$. These estimates assume
that the overall error budget is not dominated by systematic uncertainties.
With one year of SNAP data, $w_0$ could be within 10 \% provided that
the equation of state is assumed to be constant, $w=w_0$.

It is important to realise that data at low as well as high redshift
is required for an optimal parameter estimation. Events at very low
redshift help to fix the intercept ${\cal M}$, while a wide range of redshifts
is needed to break the degeneracy in the luminosity distance between
different cosmologies.

\section*{Acknowledgements}

We thank Lars Bergstr\"om, Ram Brustein, Robert Cousins, Joakim Edsj\"o, 
Antoine Letessier-Selvon, Jean-Michel Levy, 
Christian Walck and Hans-Olov Zetterstr\"om for helpful discussions. 
MG was financed by Centre National de la Recherche Scientifique (CNRS), 
France, while this work was carried out. AG is a Royal Swedish Academy 
Research Fellow supported by a 
grant from the Knut and Alice Wallenberg Foundation.

\appendix

\section{Methodology}\label{app:method}

We determine two-dimensional confidence regions
for subsets $(\theta_1,\theta_2)\in\vec{\theta}$ of the parameters 
$\vec{\theta}=(\Omega_m,\Omega_X,w_0,w_1)$, while imposing various 
conditions on the
remaining parameters. To this end, we construct log-likelihood
functions $\chi^2$ based on hypothetical
magnitude measurements at various redshifts:
\begin{equation}\label{eq:chi2-general}
  \chi^2=\sum_{i=1}^n\frac{\left[
    m(\vec{\theta}           ,{\cal M}           ,z_i)-
    m(\vec{\theta}_{\rm true},{\cal M}_{\rm true},z_i)
  \right]^2}
  {\sigma_i^2},
\end{equation}
where $m(\vec{\theta}, {\cal M},z)$ is the apparent magnitude of a supernova at
redshift $z$ in the cosmology $\vec{\theta}$ (see section \ref{sec:dl}
above), and the sum is
over bins at different redshifts. The subscript $_{\rm true}$ denotes 
actual cosmological parameter values.
The precision $\sigma_i$ of each bin is given by the individual measurement 
precision $\Delta m$ and the number of supernovae $n_i$ in the bin by 
$\sigma_i=\Delta m/\sqrt{n_i}$.

Often, we will impose prior knowledge of $\Omega_m$ and/or
$\Omega_{\rm tot}=\Omega_m+\Omega_X$. When the parameter $\theta$
of which we have prior knowledge is one of the two we are interested in,
$\theta\in(\theta_1,\theta_2)$, a Gaussian prior knowledge of $\theta$
with spread $\sigma_{\theta{\rm-prior}}$ is easily added:
\begin{equation}
  \chi^2=\chi^2_0+
  \frac{(\theta-\theta_{\rm true})^2}{\sigma_{\theta{\rm-prior}}^2},
\end{equation}
where $\chi^2_0$ denotes the $\chi^2$ obtained without imposing the
prior knowledge of $\theta$.
In case $\theta\notin(\theta_1,\theta_2)$, we have to integrate out
$\theta$ from the likelihood $L=\exp(-\frac{1}{2}\chi^2)$ with some prior
$\pi(\theta)$ to obtain $\chi^2_{\theta{\rm-int}}$:
\begin{equation}\label{eq:chi2-int}
  \chi^2_{\theta{\rm-int}}=
  -2\ln\left[\int_{-\infty}^\infty\,d\theta
  \exp\left(-\frac{1}{2}\chi^2\right)\,
  \pi(\theta)\right].
\end{equation}
Note that the form of (\ref{eq:chi2-int}) implies that a constant additive to 
$\chi^2$ simply adds to the integrated log-likelihood $\chi^2_{\theta{\rm-int}}$:
\begin{equation}
  -2\ln\left[\int\,d\theta
    \exp\left(-\frac{1}{2}(\chi^2+A)\right)\pi(\theta)\right]
    =\chi^2_{\theta{\rm-int}}+A,
\end{equation}
and that $\chi^2_{\theta{\rm-int}}-\chi^2_{\theta{\rm-int,min}}$ is
unaffected by any such constant. Consequently, we can equally well define
\begin{equation}\label{eq:chi2-int2}
  \chi^2_{\theta{\rm-int}}\equiv    
    -2\ln\left[\int\,d\theta
    \exp\left(-\frac{1}{2}(\chi^2-\chi^2_{\rm min})\right)\,\pi(\theta)\right].
\end{equation}

We will use Gaussian priors
\begin{equation}
  \pi(\theta)=\frac{1}{\sqrt{2\pi\sigma_{\theta{\rm-prior}}^2}}
  \exp\left[-\frac{1}{2\sigma_{\theta{\rm-prior}}^2}
  (\theta-\theta_{\rm true})^2\right],
\end{equation}
but also uniform priors $\pi(\theta)=1$ with $\theta$ confined to an interval
$\theta\in\theta_{\rm true}\pm\Delta\theta$.
A special case is the treatment of the intercept
${\cal M}$, for which we assume both exact knowledge, but also no
prior knowledge at all. Hence, integrating ${\cal M}$ over all possible
values ${\cal M}\in(-\infty,\infty)$, we obtain an analytic expression
for $\chi^2_{{\cal M}{\rm -int}}$, see appendix \ref{app:Mscript-int}.

Given the appropriate $\chi^2$ function, 68.3 \% and 95 \% confidence
regions are defined by the conventional two-parameter $\chi^2$ levels
2.30 and 5.99, respectively. Similarly, one-parameter one- and
two-sigma levels correspond to $\chi^2=$ 1 and 4, respectively.
In some cases we need to calculate $\chi^2$ for three parameters, and
subsequently project onto the $(\theta_1,\theta_2)$ plane of
interest. This can be done by setting $\chi^2=
{\rm min}\,[\chi^2(\cdots,\theta_3)]$, where the minimisation of $\chi^2$ 
is performed with respect to variation of $\theta_3$. Confidence regions
for $(\theta_1,\theta_2)$ can then be determined using the usual
two-parameter $\chi^2$ levels.

\subsection{Integration over the intercept ${\cal M}$}\label{app:Mscript-int} 

When the intercept is assumed to be exactly known
${\cal M}={\cal M}_{\rm true}$, it will cancel in the
expression for $\chi^2$, so that we obtain the log-likelihood
$\hat{\chi}^2$ as 
\begin{eqnarray}
  \hat{\chi}^2&\equiv&\sum_{i=1}^n\frac{\Delta^2}{\sigma_i^2} , \\
  \Delta&=&5\log_{10}\left[d'_L(\vec{\theta},z_i)\right]-
  5\log_{10}\left[d'_L(\vec{\theta}_{\rm true},z_i)\right] .
\end{eqnarray}
Note that $\hat{\chi}^2_{\rm min}=\hat{\chi}^2(\vec{\theta}_{\rm true})=0$
by construction.

If no prior knowledge of 
${\cal M}$ at all is assumed, we can integrate the general $\chi^2$ function
(\ref{eq:chi2-general}) over
${\cal M}\in(-\infty,\infty)$ to obtain an analytic expression for
$\tilde{\chi}^2\equiv\chi^2_{{\cal M}{\rm-int}}$:
\begin{eqnarray}
    \tilde{\chi}^2&=&
    -2\ln\left[\int_{-\infty}^\infty\,d{\cal M}
    \exp\left(-\frac{1}{2}\chi^2\right)\right]\\
    &=&\hat{\chi}^2-\frac{B^2}{C}+\ln\left(\frac{C}{2\pi}\right),\\ 
    B&=&\sum_{i=1}^n\frac{\Delta}{\sigma_i^2} , \\
    C&=&\sum_{i=1}^n\frac{1}{\sigma_i^2} .
\end{eqnarray}
Note that this expression is independent of ${\cal M}_{\rm true}$, and
that we imposed a uniform prior $\pi({\cal M})=1$ in the integration.
It is also worth pointing out that
\begin{equation}
  \tilde{\chi}^2_{\rm min}=
  \tilde{\chi}^2(\vec{\theta}_{\rm true})=
  \ln\left(\frac{C}{2\pi}\right) .
\end{equation}
More importantly,
\begin{equation}\label{eq:inequal}
  \tilde{\chi}^2-\tilde{\chi}^2_{\rm min}=
  \hat{\chi}^2-\frac{B^2}{C} 
  \leq\hat{\chi}^2=\hat{\chi}^2-\hat{\chi}^2_{\rm min},  
\end{equation}
where the equality holds when $B=0$.
Note that this is the case not only when
$\vec{\theta}=\vec{\theta}_{\rm true}$,
but in general also on a hypersurface in parameter space. 
The inequality (\ref{eq:inequal}) ensures the intuitive notion that
$\tilde{\chi}^2-\tilde{\chi}^2_{\rm min}$ contours
always should lie outside corresponding
$\hat{\chi}^2-\hat{\chi}^2_{\rm min}$ contours.

\subsection{Fisher matrix analysis}\label{app:fisher}

For efficient estimators (i.e., in the large sample limit), we can obtain
the Fisher matrix by finite-difference evaluation of the expression 
\begin{equation}
  F_{jk}=-\left.\frac{\partial^2\log(L)}
  {\partial\theta_j\partial\theta_k}\right|_{\vec{\theta}=\hat{\vec{\theta}}} 
  =\frac{1}{2}\left.\frac{\partial^2\chi^2}
  {\partial\theta_j\partial\theta_k}\right|_{\vec{\theta}=\hat{\vec{\theta}}}, 
\end{equation}
where, with negligible bias, we can take
$\hat{\vec{\theta}}=\vec{\theta}_{\rm true}$.
The covariance matrix is now given by the inverse of $F$.

In the quadratic approximation of $\chi^2$ (with $\chi^2$ based on the 
luminosity distance $d_L$, rather than the apparent magnitude $m$), the 
Fisher matrix is obtained as
\begin{eqnarray}
  F_{jk}&=&\sum_i h_j(z_i)\,h^{\rm T}_k(z_i) ,\label{eq:F}\\
  h_j(z_i)&=&\frac{1}{\sigma_i}
  \left.\frac{\partial d_L}{\partial \theta_j}
  \right|_{\vec{\theta}=\hat{\vec{\theta}};z=z_i} ,
\end{eqnarray}
where the precision can be expressed in terms of the relative precision $p$ 
as $\sigma_i=p\,d_L(z_i)$. It is straight-forward to add prior knowledge of 
any combination of
the parameters $\vec{\theta}$. Imposing no prior knowledge of
${\cal M}$ corresponds to letting the scale of $d_L$ be unknown:
$d_L=Q\,d_L'$. 

It should be noted that, even though equation (\ref{eq:F}) 
is an approximation, it gives uncertainties in accordance with the analysis 
in section \ref{sec:snap} (compare, for instance, maximum values in 
figures \ref{fig:add01-06} -- \ref{fig:add31-36} with relevant cases in 
tables \ref{tab:sigma1} and \ref{tab:sigma3}). In addition, 
for inefficient estimators (i.e., non-ellipsoidal confidence regions), 
the approximate Fisher analysis roughly gives the mean errors of parameters.

\onecolumn

\begin{table}
  \centering
  \begin{tabular}{llcc}
    & & $\sigma_{\Omega_m}$ & $\sigma_{\Omega_X}$ \\ \hline
    $(\Omega_m,\Omega_X)$
    & exact ${\cal M}$, no prior $\Omega_m$
    & $\pm0.015$ & $\pm0.027$ \\
    & exact ${\cal M}$, Gaussian $\Omega_m$, 
    $\sigma_{\Omega_m{\rm-prior}}=0.05$
    & $\pm0.015$ & $\pm0.026$ \\
    & no prior ${\cal M}$, no prior $\Omega_m$
    & $\pm0.017$ & $\pm0.047$ \\
    & no prior ${\cal M}$, Gaussian $\Omega_m$, 
    $\sigma_{\Omega_m{\rm-prior}}=0.05$
    & $\pm0.016$ & $\pm0.045$ \\ \hline
    $(\Omega_m,\Omega_X)$ (no $z\in[1.2,1.7]$ events)
    & exact ${\cal M}$, no prior $\Omega_m$
    & $\pm0.020$ & $\pm0.033$ \\
    & exact ${\cal M}$, Gaussian $\Omega_m$, 
    $\sigma_{\Omega_m{\rm-prior}}=0.05$
    & $\pm0.019$ & $\pm0.031$ \\
    & no prior ${\cal M}$, no prior $\Omega_m$
    & $\pm0.024$ & $\pm0.058$ \\
    & no prior ${\cal M}$, Gaussian $\Omega_m$, 
    $\sigma_{\Omega_m{\rm-prior}}=0.05$
    & $\pm0.021$ & $\pm0.053$ \\ \hline
    $(\Omega_m,\Omega_X)$ (constant rate/volume at $z\in[0,1.2]$)
    & exact ${\cal M}$, no prior $\Omega_m$
    & $\pm0.016$ & $\pm0.030$ \\
    & exact ${\cal M}$, Gaussian $\Omega_m$, 
    $\sigma_{\Omega_m{\rm-prior}}=0.05$
    & $\pm0.015$ & $\pm0.028$ \\
    & no prior ${\cal M}$, no prior $\Omega_m$
    & $\pm0.017$ & ${}^{+0.079}_{-0.087}$ \\
    & no prior ${\cal M}$, Gaussian $\Omega_m$, 
    $\sigma_{\Omega_m{\rm-prior}}=0.05$
    & $\pm0.016$ & ${}^{+0.077}_{-0.084}$ \\ \hline
  \end{tabular}
  \caption{One-parameter one-sigma ranges for $(\Omega_m,\Omega_X)$ 
    in the one-year SNAP scenario. The quoted parameter ranges for a 
    parameter $\theta$ are obtained by finding the extremal values of 
    $\theta$ for which $\chi^2=1$.}
  \label{tab:sigma1} 
\end{table}

\begin{table}
  \centering
  \begin{tabular}{llccc}
    & & $\sigma_{\Omega_m}$ & $\sigma_{\Omega_X}$ & $\sigma_{w_0}$ \\ \hline
    $(\Omega_m,w_0)$, fixed $\Omega_{\rm tot}=1$
    & exact ${\cal M}$, no prior $\Omega_m$
    & ${}^{+0.003}_{-0.019}$ & -- & $+0.048$ \\
    & exact ${\cal M}$, Gaussian $\Omega_m$, 
    $\sigma_{\Omega_m{\rm-prior}}=0.05$
    & ${}^{+0.003}_{-0.017}$ & -- & $+0.045$ \\
    & no prior ${\cal M}$, no prior $\Omega_m$
    & ${}^{+0.007}_{-0.023}$ & -- & $+0.078$ \\
    & no prior ${\cal M}$, Gaussian $\Omega_m$, 
    $\sigma_{\Omega_m{\rm-prior}}=0.05$
    & ${}^{+0.007}_{-0.021}$ & -- & $+0.071$ \\ \hline
    $(\Omega_m,w_0)$, fixed $\Omega_X=0.72$
    & exact ${\cal M}$, no prior $\Omega_m$
    & ${}^{+0.006}_{-0.017}$ & -- & $+0.030$ \\
    & exact ${\cal M}$, Gaussian $\Omega_m$, 
    $\sigma_{\Omega_m{\rm-prior}}=0.05$
    & ${}^{+0.006}_{-0.016}$ & -- & $+0.028$ \\
    & no prior ${\cal M}$, no prior $\Omega_m$
    & ${}^{+0.010}_{-0.021}$ & -- & $+0.055$ \\
    & no prior ${\cal M}$, Gaussian $\Omega_m$, 
    $\sigma_{\Omega_m{\rm-prior}}=0.05$
    & ${}^{+0.010}_{-0.019}$ & -- & $+0.052$ \\ \hline
    $(\Omega_X,w_0)$, fixed $\Omega_m=0.28$
    & exact ${\cal M}$, no prior $\Omega_X$
    & -- & ${}^{+0.15}_{-0.010}$ & $+0.12$ \\
    & exact ${\cal M}$, Gaussian $\Omega_X$, 
    $\sigma_{\Omega_X{\rm-prior}}=0.05$
    & -- & ${}^{+0.048}_{-0.010}$ & $+0.045$ \\
    & no prior ${\cal M}$, no prior $\Omega_X$
    & -- & ${}^{+0.15}_{-0.026}$ & $+0.12$ \\
    & no prior ${\cal M}$, Gaussian $\Omega_X$, 
    $\sigma_{\Omega_X{\rm-prior}}=0.05$
    & -- & ${}^{+0.048}_{-0.024}$ & $+0.049$ \\ \hline
    \multicolumn{2}{l}{$(\Omega_X,w_0)$, 
      Gaussian $\Omega_m$, $\sigma_{\Omega_m{\rm-prior}}=0.05$
      and Gaussian $\Omega_{\rm tot}$, 
      $\sigma_{\Omega_{\rm tot}{\rm-prior}}=0.05$,
      exact ${\cal M}$}
    & -- & ${}^{+0.06}_{-0.02}$ & $+0.07$ \\ 
    \multicolumn{2}{l}{$(\Omega_X,w_0)$, 
      Gaussian $\Omega_m$, $\sigma_{\Omega_m{\rm-prior}}=0.05$
      and Gaussian $\Omega_{\rm tot}$, 
      $\sigma_{\Omega_{\rm tot}{\rm-prior}}=0.05$,
      no prior ${\cal M}$}
    & -- & ${}^{+0.06}_{-0.03}$ & $+0.09$ \\ \hline
  \end{tabular}
  \caption{One-parameter one-sigma ranges for $(\Omega_m,w_0)$ or
    $(\Omega_X,w_0)$ in the one-year SNAP scenario. 
    The quoted parameter ranges for a parameter $\theta$ are 
    obtained by finding the extremal values of $\theta$ for which 
    $\chi^2=1$, with the additional requirement $w_0\ge-1$.}
  \label{tab:sigma2} 
\end{table}

\begin{table}
  \centering
  \begin{tabular}{llcc}
    & & $\sigma_{w_0}$ & $\sigma_{w_1}$ \\ \hline
    $(w_0,w_1)$, $\Omega_{\rm tot}=1$
    & exact ${\cal M}$, exact $\Omega_m$ 
    & $\pm0.031$ & $\pm0.14$ \\
    & exact ${\cal M}$, $\Omega_m\in\Omega_{m,{\rm true}}\pm0.1$
    & ${}^{+0.13}_{-0.066}$ & ${}^{+0.48}_{-0.76}$ \\
    & exact ${\cal M}$, Gaussian $\Omega_m$, 
    $\sigma_{\Omega_m{\rm-prior}}=0.05$ 
    & ${}^{+0.065}_{-0.052}$ & ${}^{+0.31}_{-0.46}$ \\
    & no prior ${\cal M}$, exact $\Omega_m$
    & $\pm0.064$ & $\pm0.18$ \\ 
    & no prior ${\cal M}$, Gaussian $\Omega_m$, 
    $\sigma_{\Omega_m{\rm-prior}}=0.05$ 
    & ${}^{+0.077}_{-0.074}$ & ${}^{+0.35}_{-0.53}$ \\ \hline
    $(w_0,w_1)$, (no $z\in[1.2,1.7]$ events)
    & exact ${\cal M}$, exact $\Omega_m$ 
    & $\pm0.034$ & $\pm0.16$ \\
    & exact ${\cal M}$, $\Omega_m\in\Omega_{m,{\rm true}}\pm0.1$
    & ${}^{+0.13}_{-0.085}$ & ${}^{+0.50}_{-1.03}$ \\
    & exact ${\cal M}$, Gaussian $\Omega_m$, 
    $\sigma_{\Omega_m{\rm-prior}}=0.05$ 
    & ${}^{+0.068}_{-0.059}$ & ${}^{+0.32}_{-0.48}$ \\
    & no prior ${\cal M}$, exact $\Omega_m$
    & $\pm0.070$ & $\pm0.21$ \\ 
    & no prior ${\cal M}$, Gaussian $\Omega_m$, 
    $\sigma_{\Omega_m{\rm-prior}}=0.05$ 
    & ${}^{+0.085}_{-0.083}$ & ${}^{+0.36}_{-0.54}$ \\ \hline
    $(w_0,w_1)$, (constant rate/volume at $z\in[0,1.2]$)
    & exact ${\cal M}$, exact $\Omega_m$ 
    & $\pm0.038$ & $\pm0.16$ \\
    & exact ${\cal M}$, $\Omega_m\in\Omega_{m,{\rm true}}\pm0.1$
    & ${}^{+0.13}_{-0.066}$ & ${}^{+0.50}_{-0.96}$ \\
    & exact ${\cal M}$, Gaussian $\Omega_m$, 
    $\sigma_{\Omega_m{\rm-prior}}=0.05$ 
    & ${}^{+0.065}_{-0.054}$ & ${}^{+0.33}_{-0.53}$ \\
    & no prior ${\cal M}$, exact $\Omega_m$
    & ${}^{+0.14}_{-0.15}$ & ${}^{+0.27}_{-0.26}$ \\ 
    & no prior ${\cal M}$, Gaussian $\Omega_m$, 
    $\sigma_{\Omega_m{\rm-prior}}=0.05$ 
    & ${}^{+0.14}_{-0.15}$ & ${}^{+0.42}_{-0.66}$ \\ \hline
    $(w_0,w_1)$ (Maor et al. cosmology)
    & exact ${\cal M}$, exact $\Omega_m$
    & $\pm0.028$ & $\pm0.11$ \\
    & exact ${\cal M}$, $\Omega_m\in\Omega_{m,{\rm true}}\pm0.1$
    & ${}^{+0.10}_{-0.10}$ & ${}^{+0.26}_{-0.63}$ \\
    & exact ${\cal M}$, Gaussian $\Omega_m$, 
    $\sigma_{\Omega_m{\rm-prior}}=0.05$ 
    & ${}^{+0.054}_{-0.052}$ & ${}^{+0.16}_{-0.22}$ \\
    & no prior ${\cal M}$, exact $\Omega_m$
    & $\pm0.057$ & $\pm0.16$ \\
    & no prior ${\cal M}$, Gaussian $\Omega_m$, 
    $\sigma_{\Omega_m{\rm-prior}}=0.05$ 
    & ${}^{+0.070}_{-0.072}$ & ${}^{+0.20}_{-0.25}$ \\ \hline\hline
    $(w_0,w_1)$, $\Omega_{\rm tot}=1$, (three-year SNAP)
    & exact ${\cal M}$, exact $\Omega_m$ 
    & $\pm0.018$ & $\pm0.081$ \\
    & exact ${\cal M}$, Gaussian $\Omega_m$, 
    $\sigma_{\Omega_m{\rm-prior}}=0.05$ 
    & ${}^{+0.060}_{-0.038}$ & ${}^{+0.29}_{-0.36}$ \\
    & exact ${\cal M}$, Gaussian $\Omega_m$, 
    $\sigma_{\Omega_m{\rm-prior}}=0.015$ 
    & ${}^{+0.023}_{-0.024}$ & ${}^{+0.13}_{-0.15}$ \\
    & no prior ${\cal M}$, exact $\Omega_m$
    & $\pm0.036$ & $\pm0.11$ \\ 
    & no prior ${\cal M}$, Gaussian $\Omega_m$, 
    $\sigma_{\Omega_m{\rm-prior}}=0.05$ 
    & ${}^{+0.062}_{-0.047}$ & ${}^{+0.31}_{-0.42}$ \\
    & no prior ${\cal M}$, Gaussian $\Omega_m$, 
    $\sigma_{\Omega_m{\rm-prior}}=0.015$ 
    & ${}^{+0.038}_{-0.039}$ & ${}^{+0.15}_{-0.17}$ \\ \hline\hline
    $(w_0,w_1)$ (Maor et al. scenario)
    & exact ${\cal M}$, exact $\Omega_m$
    & $\pm0.014$ & $\pm0.044$ \\
    & exact ${\cal M}$, $\Omega_m\in\Omega_{m,{\rm true}}\pm0.1$
    & ${}^{+0.094}_{-0.051}$ & ${}^{+0.21}_{-0.34}$
  \end{tabular}
  \caption{One-parameter one-sigma ranges for $(w_0,w_1)$ in the
    one-year SNAP scenario. 
    Note that the two last sections instead refer to the three-year SNAP scenario
    and the scenario of Maor et al. (\protect\cite{maor}), respectively,
    also discussed in section \ref{sec:conf-w0-w1}. 
    The quoted parameter ranges for a parameter $\theta$ are obtained by 
    finding the extremal values of $\theta$ for which $\chi^2=1$.}
  \label{tab:sigma3} 
\end{table}

\end{document}